\documentstyle[psfig,12pt]{ioplppt}
\newcommand{\nc}{\newcommand}
\newcommand{\rnc}{\renewcommand}
\nc{\inodot}{\char'020}
\rnc{\/}{\over}
\nc{\<}{\langle}
\rnc{\>}{\rangle}
\rnc{\(}{\left (}
\rnc{\)}{\right )}
\rnc{\[}{\left [}
\rnc{\]}{\right ]}
\rnc{\i}{{\rm i}}
\nc{\bm}[1]{{\mbox{\boldmath$#1$}}}
\nc{\applss}{\,\lower0.5ex\hbox{$\sim$}\kern-0.79em\raise0.5ex\hbox{$<$}\,}
\nc{\appgtr}{\,\lower0.5ex\hbox{$\sim$}\kern-0.79em\raise0.5ex\hbox{$>$}\,}
\nc{\figdir}{EPS}
\begin{document}
\jl{1}

\title{Spectral Statistics in Chaotic Systems with Two Identical, Connected
Cells}[Chaotic Systems with Two Identical Cells]

\author{
 T.~Dittrich$^{1}$\footnote{
  Permanent address: Depto. de F\'\inodot sica, Universidad de los Andes,
  A.~A.~4976, Santaf\'e de Bogot\'a, Colombia
 },
 G.~Koboldt$^{2}$\footnote{
  Present address: Fachbereich 7 -- Physik,
  Universit\"at GHS Essen, D-45117 Essen, Germany
 }, 
 B.~Mehlig$^{1}$, 
 H.~Schanz$^{3}$
}

\address{$^{1}$Max-Planck-Institut f\"ur Physik komplexer Systeme, 
N\"othnitzer Stra\ss e 38,\\ D-01187 Dresden, Germany}
\vspace*{2mm}

\address{$^{2}$Institut f\"ur Physik, Universit\"at Augsburg, Memminger Stra\ss e 6,\\
D-86135 Augsburg, Germany}
\vspace*{2mm}

\address{$^{3}$Max-Planck-Institut f\"ur Str\"omungsforschung,
Bunsenstra\ss e 10,\\ D-37073 G\"ottingen, Germany}

\date{July 9, 1998}

\begin{abstract}
Chaotic systems that decompose into two cells connected only by a narrow
channel exhibit characteristic deviations of their quantum spectral statistics
from the canonical random-matrix ensembles. The equilibration between the
cells introduces an additional classical time scale that is manifest also in
the spectral form factor. If the two cells are related by a spatial symmetry,
the spectrum shows doublets, reflected in the form factor as a positive peak
around the Heisenberg time. We combine a semiclassical analysis with an
independent random-matrix approach to the doublet splittings to obtain the
form factor on all time (energy) scales. Its only free parameter is the
characteristic time of exchange between the cells in units of the Heisenberg
time.
\end{abstract}

\maketitle

\section{Introduction}
\label{sec:intro}

Most of the pioneering enquiries into quantum chaos have been focussed on
bounded systems---closed billiards, atomic systems---whose classical dynamics
knows only a single global time scale, the inverse Kolmogorov entropy which
describes the ergodic coverage of phase space. The absence of other classical
times has facilitated the understanding of quantum-to-classical relationships
and has opened the view for universal features in spectra and eigenfunctions
of classically chaotic systems.

Extended systems represent another, still simple, extreme. They reach
ergodicity only on a time scale that exceeds all other characteristic
times. With extended systems, basic solid-state concepts enter quantum
chaos. The crucial r\^ole of long-range spatial order, in particular, for
spectrum and transport has to be considered in the unfamiliar context of
dynamical disorder. In a semiclassical approach, some light could recently be
shed on the spectral signatures of chaotic diffusion, both in the band
structure of periodic systems \cite{dit1} and in the discrete spectra of
disordered systems with localized eigenstates \cite{dit2,dit3,arg}.

\begin{figure}
\centerline{\psfig{figure=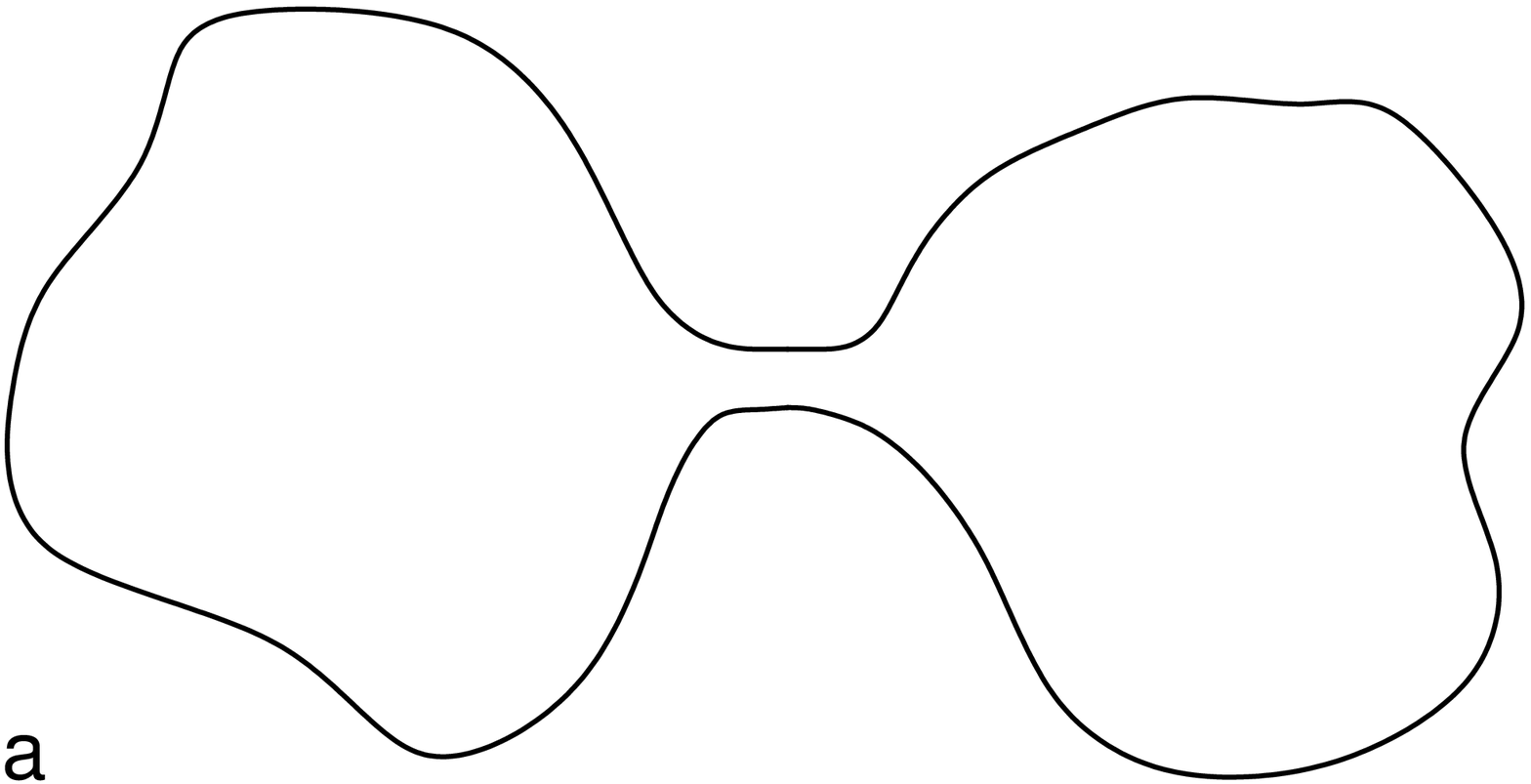,width=6cm}
\psfig{figure=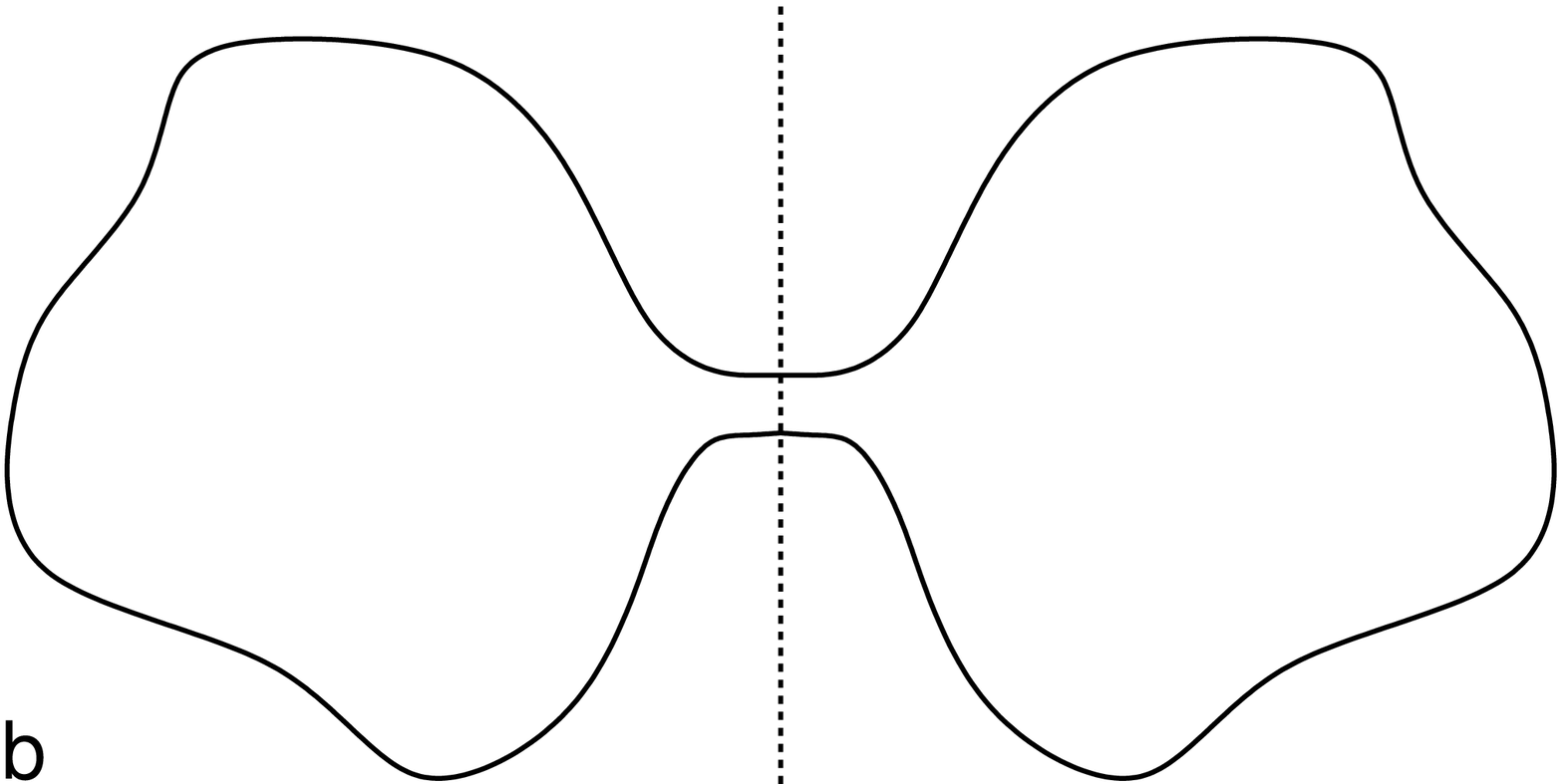,width=6cm}}
\caption{\label{bones} 
Two-cell billiard with narrow bottleneck, of arbitrary (a) and
reflection-symmetric shape (b).}
\end{figure}

A region intermediate between bound and extended is marked by systems
comprising just a few weakly connected similar cells. In the following, we
will consider spatial confinements or phase-space structures that decompose
into two compartments, connected only by a ``bottleneck'' (Fig.~\ref{bones})
\cite{smi}. Equilibration between the cells then takes much longer than the
ergodic coverage of a single cell. It constitutes a second independent time
scale of the classical dynamics.

Chaotic systems with two cells have quite diverse applications. The phase space
of alkali atoms in a strong magnetic field may contain two almost disjunct
chaotic regions, related by a spatial symmetry \cite{cam}. Similar systems arise
as models for non-ergodic reaction dynamics in quantum chemistry \cite{del}. 
More generally, they form prototypes of systems decomposing into two similar,
weakly coupled parts. In this sense, they can represent heavy nuclei in a final
state of fission \cite{bra}, or nuclei with their two isospin subsystems
interacting weakly due to a slight breaking of isospin invariance \cite{ric}.

In their quantum-mechanical properties, two-cell systems already exhibit the
decisive influence of spatial symmetry. If the two cells are of arbitrary
shape or size (Fig.~\ref{bones}a), their restricted communication will merely
be reflected in a quantitative deviation from the canonical random-matrix
statistics \cite{smi}. It does not introduce any qualitatively new feature.
The situation changes considerably if the cells are related by some twofold
spatial symmetry (Fig.~\ref{bones}b). A genuine quantum phenomenon, a coherent
mode of transport between the cells emerges, and the spectrum shows systematic
quasidegeneracies. This clustering of levels is manifest in the two-point
correlations as a marked {\em positive} peak on the scale of the mean
single-cell level separation, or in the time domain, the single-cell
Heisenberg time \cite{dit1}.

This phenomenon should be carefully distinguished from tunneling. To be sure,
the doublets do resemble tunnel splittings in the sense that they are based on
a discrete spatial symmetry and correspond to quantum coherent transport on
very long time scales. Moreover, in the wavenumber regime where there is no
open channel in the constriction between the cells, the wavefunctions decay
exponentially into this region. Even at higher wavenumber, the amplitude is
often strongly suppressed there (cf.\ Fig.~\ref{cont-fig}). Since, however,
there is neither a potential nor a dynamical barrier involved, this transport
is slow but not classically forbidden.

Chaos-assisted tunneling, in contrast, is a hallmark of bistable systems with a
mixed phase space \cite{boh1,tom,ute}. It occurs between symmetry-related pairs
of {\em regular} islands in phase space that are separated by a chaotic region.
Here, by contrast, we are dealing with symmetry-related {\em chaotic} regions
communicating through a narrow bridge in space. Still, a similar situation can
occur also in mixed bistable systems. In fact, the distribution of the
splittings of doublet states supported by symmetry-related pairs of chaotic
regions \cite{cre} forms an important input to the distribution of tunnel
splittings in chaos-assisted tunneling \cite{ley}.

In the following sections, we shall develop a theory for the spectral
statistics of two-cell systems that largely rests on recent progress in the
analysis of band structures of classically chaotic systems with spatial
periodicity \cite{dit1}. There, an important input has been the notion of form
factors with a winding-number argument, specific for transitions spanning a
corresponding number of unit cells. Likewise, the group property of a
reflection or translation symmetry of a two-cell system enables the definition
of form factors with a rudimentary spatial resolution, expressed by a binary
index that indicates either return to the same cell or transport into the
opposite cell.

However, as compared to \cite{dit1}, we do not only go from a large value of
$N$, the number of unit cells, to $N = 2$. Here, we shall concentrate on the
case of slow exchange between the cells, the ``weak-coupling'' or
``tight-binding'' limit in solid-state terminology, while \cite{dit1} was
devoted to the opposite case. This implies that, even if the two-cell system
is ``unfolded'' to form an infinite chain (see \ref{app:dif}), the
concept of homogeneous diffusion no longer applies. Rather, in the unfolded
picture, we are dealing with a spatially discrete diffusion process that
deviates significantly, on short time scales, from ordinary
diffusion. Concerning the spectral statistics, the principal consequence is
that we are now dealing with {\em flat} ``bands''.

In order to extend a semiclassical treatment of the spectral correlations to
energy scales below the mean level spacing, information on the spreading of
the possibly complex trajectories that mediate the long-time transitions
\cite{leb,cre}, analogous to the sum rules for ergodicity \cite{han} or
diffusion \cite{dit2,dit3}, would be required. Alternatively, we would have to
surmount the diagonal approximation. We circumvent this open problem and
instead adopt a different strategy. In the spirit of Berry's semiclassical
approximation for the form factors of random-matrix ensembles \cite{ber}, we
impose plausible assumptions on the distribution of the narrow splittings.
They are based on the relation of the doublet splittings in two-cell compounds
to the resonance widths in corresponding single-cell systems, obtained by
opening up the system at the constriction. We derive this relation in the case
of a single open channel between the cells, where doublet splittings and
resonance widths are not simply identical.

Switching back from the distribution of doublet splittings to the
corresponding time-domain function valid on long time scales, we match the
resulting long-time asymptote with the semiclassical short-time behaviour. In
this way, we achieve expressions for the spectral two-point correlations on
all time scales. They are universal in that they contain, as the only free
parameter, the characteristic time for equilibration between the cells. It is
the two-cell analogue of the conductance, the scaling parameter in the case of
long chains.

We shall introduce the classical concepts relevant for the dynamics of systems
with two identical, connected cells in the subsequent Section
\ref{sec:cls}. In Section \ref{sec:fof}, we define form factors specific for
an element of the symmetry group of the system. A semiclassical theory for the
short-time regime of these form factors is developed in Section \ref{sec:scl},
while Section \ref{sec:qum} is devoted to their quantum long-time
behaviour. Some of these calculations are extended to the case of unrestricted
values of $N$ in two appendices.  Section \ref{sec:num} serves to introduce
four illustrative models, two versions of a Sinai billiard \cite{sin,sch},
quantum graphs \cite{kot} configured in such a way that they form a two-cell
system, a two-cell variant of the quantum kicked rotor \cite{izr,dit1}, and a
random-matrix model \cite{boh3,tom}. Spectral data obtained numerically for these
models corroborate our theory. Section \ref{sec:sum} contains a synopsis of the
various limiting cases covered in this paper.

\section{Classical dynamics in two-cell systems}
\label{sec:cls}
As a minimal version of a classical two-cell system, consider the following
model: Two spatially confined compartments are connected by some narrow duct
(Fig.~\ref{bones}a). In view of the intended applications, we require a few
additional properties. The leakage from the cell where the system is prepared
(subscript `0' in the following) to the opposite side (subscript `1') should
be completely described by a single time scale $1/\lambda$. This amounts to an
exponential decay of the population from the initial cell, if it were opened
by removing the opposite cell. We require the rate $\lambda$ to be the same in
both directions.  A sufficient condition for this to be true is that the cells
form a (translation- or reflection-) symmetric pair
(Fig.~\ref{bones}b). Finally, we assume that the dynamics within the cells is
chaotic and thus ergodic, and that coverage of a single cell is reached
instantaneously on the scale $1/\lambda$.

Under these conditions, the time evolution of the probability to stay in
either cell obeys the simple pair of master equations,
\begin{eqnarray}\label{mastereq}
\begin{array} {rcl}
{\displaystyle \dot{P}_0(t)} &=&
{\displaystyle \lambda\left(P_1(t) - P_0(t)\right),} \\*[0.2cm]
{\displaystyle \dot{P}_1(t)} &=&
{\displaystyle \lambda\left(P_0(t) - P_1(t)\right).}
\end{array}
\end{eqnarray}
The relaxation into equilibrium of the two probabilities, given by
$\lim_{t\to\infty} P_{0/1}(t)/(P_0(t)+P_1(t)) = 1/2$, is governed by the rate
$\Lambda = 2\lambda$. From an initial state $P_0(0) = 1$, $P_1(0) = 0$, they
evolve as
\begin{equation}\label{mastersol}
P_{0/1}(t) = \frac{1}{2} (1 \pm {\rm e}^{-\Lambda t}).
\end{equation}
The population difference
\begin{equation}
P_{\rm d}(t) = P_0(t) - P_1(t)
\end{equation}
is another relevant quantity. Besides the sum $P_0(t) + P_1(t)$, it plays the
r\^ole of an eigenmode amplitude of the master equation (\ref{mastereq}). Its
time evolution reads, for the same initial state as above,
\begin{equation}
P_{\rm d}(t) = {\rm e}^{-\Lambda t}.
\end{equation}
The above considerations apply also if the cells communicate through two or
more physical channels. This includes in particular the case of two cells
connected at both ``ends'' to form a ring. The rates of probability exchange
through the channels then just add to give the global rate $\lambda$. The
diffusive dynamics that results if a two-cell ring configuration is unrolled
into an infinite chain, is discussed in \ref{app:dif}.

As an example, we state the explicit expression for the decay rate $\lambda$
in the case of an ergodic double billiard as in Fig.~\ref{bones}. The
phase-space area leaving one cell of the billiard in time ${\rm d}t$ through the
connecting channel of width $s$, at unit speed, is ${\rm d}\Omega = 2s{\rm d}t$.
This is to be normalized by the area $A$ of a cell and by $2\pi$, the size of
momentum space projected onto the energy shell. The resulting expression for the
escape rate is
\begin{equation}
\label{rate}
\lambda_{\rm erg} = \frac{s}{\pi A}.
\end{equation}

\section{Generalized form factors}
\label{sec:fof}

In quantum systems, each unitary symmetry gives rise to a constant of the
motion, a ``good quantum number''. It takes as many values as there are
irreducible re\-pre\-sen\-tations of the symmetry, and the full spectrum can
be decomposed into subspectra, each of which pertains to a given irreducible
representation. Formally, the decomposition is effected by the projectors
\cite{tin} $\widehat P_\nu = N^{-1} \sum_{n=0}^{N-1} \chi_\nu(g_n) \widehat
U^{\dagger}(g_n)$, $\nu = 0$, $\ldots$, $N-1$. Here, $N$ is the number of
elements $g_n$ of the symmetry group and simultaneously the number of its
representations (for simplicity, we assume all representations to be one
dimensional). The character of $g_n$ in the $\nu$th representation is referred
to as $\chi_\nu(g_n)$, and $\widehat U(g_n)$ denotes the unitary
transformation corresponding to $g_n$. Spectral densities and correlation
functions within a given representation can then be defined on basis of the
symmetry-projected Green function $\widehat G_\nu(E) = \widehat P_\nu \widehat
G(E)$, with $\widehat G(E)$, the Green function for the entire spectrum. For
example, the symmetry-projected spectral density in the $\nu$th representation
is defined as
\begin{equation}
\label{symredden}
\tilde d_\nu(E) = \sum_\alpha \delta(E - E_{\alpha,\nu})
= -\frac{1}{\pi}{\rm Im}\,{\rm tr}\,[\widehat G_\nu(E)],
\end{equation}
where the $E_{\alpha,\nu}$ are the eigenenergies in the $\nu$th representation.

For the study of quasidegenerate doublets, another type of spectral density is
as relevant as the symmetry-projected one. As an alternative to $\tilde
d_\nu(E)$, one may define densities and derived quantities that refer to a
group element $g_n$, instead of an irreducible representation $\nu$. The
symmetry group induces a tiling of (phase) space, i.e., a decomposition into
disjunct segments such that each of them is mapped onto all the others by the
transformations in the group, thus covering the entire space \cite{rob}. A
group-element-specific spectral density therefore provides a rudimentary spatial
resolution on the scale of the fundamental domain of the group.

If all representations are one dimensional, the set of columns of the matrix
$\chi_{\nu,n} = \chi_\nu(g_n)$ of group characters forms an orthogonal basis in
$N$ dimensions \cite{tin}. The same is true for the rows. The matrix as a
whole therefore has full rank and is invertible. We refer to the inverse
matrix as $\chi^{-1}$. Multiplying the vector of symmetry-reduced spectral
densities $\tilde d_\nu(E)$, Eq.~(\ref{symredden}), from the left by
$\chi^{-1}$, we obtain the spectral densities \cite{dit1}
\begin{eqnarray}
d_n(E) &=& \sum_{\nu=0}^{N-1}
(\chi^{-1})_{\nu,n} \tilde d_\nu(E) \label{elspecden1} \\
&=& -\frac{1}{\pi}{\rm Im} \int_{\rm fd} {\rm d}q\,
\widehat G(g_n({\bm q}),{\bm q};E). \label{elspecden3}
\end{eqnarray}
The space integral in Eq.~(\ref{elspecden3}) extends only over the
fundamental domain (subscript `fd') of
the tesselated space. The first argument of the Green function in the third
line is the image of its second argument, ${\bm q}$, under $g_n$. This suggests
that $d_n(E)$ refers to transitions from any segment to its image under
$g_n$.

By Fourier transforming Eq.~(\ref{elspecden1}) with respect to energy, we
arrive at the analogous density in the time domain,
\begin{eqnarray}
a_n(\tau) &=& \< d_{\rm fd}\>^{-1}\int_{-\infty}^{\infty}
{\rm d}r\, {\rm e}^{-2\pi{\rm i}r\tau}
d_n(r/\< d_{\rm fd}\>) \label{elspecamp1} \\
&=& \int_{\rm fd} {\rm d}q\,
\< g_n({\bm q})|\widehat U(t_{\rm H}\tau)
|{\bm q}\>. \label{elspecamp2}
\end{eqnarray}
We have switched to dimensionless energy, $r = \< d_{\rm fd}\> E$,
and time, $\tau = t/t_{\rm H}$, by scaling with the mean spectral density
$\< d_{\rm fd}\>$ in the symmetry-reduced space, and the
corresponding Heisenberg time $t_{\rm H} = 2\pi\hbar\< d_{\rm
fd}\>$, respectively. Equation (\ref{elspecamp2}) describes an amplitude
to return modulo the symmetry transformation $g_n$. The corresponding return
probability is given by the form factor
\begin{equation}
\label{formfacdef}
K_n(\tau) = \frac{1}{\Delta r_{\rm fd}} |a_n(\tau)|^2\,,
\end{equation}
where $\Delta r_{\rm fd}$ is the total width, in units of $\< d_{\rm fd}\>$, 
of the spectrum considered. Eq.~(\ref{formfacdef}) is equivalent to a definition
of the form factor as the Fourier transform of a spectral autocorrelation
function \cite{dit1,dit3}.

For a two-cell system, a symmetry that maps one cell to the other can be a
reflection or a translation. Their group elements are identity (denoted by $n
= 0$ in the following) and reflection or translation ($n = 1$, without
distinguishing the two). The characters are both 1 in the symmetric (subscript
`$+$') representation, and $\pm 1$ in the antisymmetric (`$-$')
representation. Following the general discussion above, we define symmetric
and antisymmetric form factors, respectively, by
\begin{equation}
\label{saformfac}
\widetilde K_{\pm}(\tau) = \frac{1}{\Delta r_{\rm fd}} 
|\tilde a_{\pm}(\tau)|^2\,.
\end{equation}
The amplitudes $\tilde a_{\pm}(\tau)$ are obtained, for example, by sorting
the spectral data of the two-cell system according to the symmetry of the
corresponding eigenstates, and Fourier transforming as in
Eq.~(\ref{elspecamp1}).  Alternatively, $\widetilde K_{\pm}$ can be
interpreted as the form factors of a single cell with Neumann or Dirichlet
boundary conditions, respectively, imposed on the line(s) in configuration
space common to the two cells. In general, this choice of boundary conditions
does not affect the chaoticity of the classical dynamics. Hence we expect the
form factor of a single cell to equal the random-matrix result \cite{boh2} up
to normalization, i.e.,
\begin{equation}\label{sff-rmt}
\widetilde K_{\pm}(\tau)={1\/2}K_{\rm RMT}(\tau)\,.
\end{equation}
Form factors specific for return to the initial (subscript `0') or switching
to the opposite cell (`1') are defined according to Eq.~(\ref{elspecden1}) and
Eqs.~(\ref{elspecamp1}), (\ref{formfacdef}) as
\begin{equation}
\label{twoelformfac}
K_{0/1}(\tau) = \frac{1}{\Delta r_{\rm fd}}
|\tilde a_+(\tau) \pm \tilde a_-(\tau)|^2\,.
\end{equation}
Note that the amplitudes are superposed {\em before} squaring. We shall
show in Sect.~\ref{sec:qum} below that considering the {\em incoherent}
superpositions $K_0(\tau) \pm K_1(\tau)$, in turn, provides an
approximate access to the distribution of doublet splittings and inter-doublet
separations, respectively.

The combination of amplitudes $\tilde a_+(\tau)+\tilde a_-(\tau)$ entering
$K_{0}(\tau)$ in Eq.~(\ref{twoelformfac}) is obtained in (\ref{elspecamp1}),
when $d_{n}$ on the r.~h.~s.\ is replaced by the total spectral density
of the two-cell system. Hence $K_{0}$ is---up to scaling of time and
energy---equivalent to a form factor defined without any reference to spatial
symmetry of the two cells.

In Eqs.~(\ref{saformfac}) and (\ref{twoelformfac}) we have chosen a
normalization which ensures that $\widetilde K_{\pm}(\tau)$ and
$K_{0/1}(\tau)$ approach the same value $1/2$ for $\tau\to\infty$, provided
the two cells are not completely disconnected. In addition, the sums of the
symmetry-projected and the group-element-specific form factors are the same
for arbitrary time $\tau$,
\begin{equation}
\label{presofnorm}
K_0(\tau) + K_1(\tau) = \widetilde K_+(\tau) + \widetilde K_-(\tau)\,.
\end{equation}
This identity may be interpreted as a preservation of norm and follows
generally from the unitarity of the matrix $\chi_{\nu,n}$ of group characters.
Because of (\ref{sff-rmt}) it leads to the relation 
\begin{equation}\label{tc-rmt}
K_0(\tau) + K_1(\tau)=K_{\rm RMT}(\tau)\,.
\end{equation}
It is instructive at this stage, to consider the trivial limiting cases of two
completely isolated or two very strongly interacting cells, respectively. In
the first case, we have from the definitions (\ref{elspecden1}),
(\ref{elspecamp1}) and (\ref{twoelformfac}) $K_{1}(\tau)=0$. Then
(\ref{sff-rmt}) implies $K_{0}(\tau)=K_{\rm RMT}(\tau)$, and this is indeed
what is expected within our scaling of time and energy from the fact that the
total spectrum is the superposition of two {\em identical} random-matrix
spectra.  In the other extreme, the two subspectra of positive and negative
parity can be considered statistically independent, which has the consequence
$K_{0}(\tau)=K_{1}(\tau)=K_{\rm RMT}(\tau)/2$.

\section{Semiclassical regime}
\label{sec:scl}
In order to construct a semiclassical trace formula for the symmetry-projected
spectral density $\tilde d_\nu(E)$, Eq.~(\ref{symredden}), the concept of
periodic orbits has to be extended \cite{rob}. In case the dynamics within the
cells has no significant admixture of regular motion, the trace formula reads
\begin{eqnarray}
\label{symprotrace}
\tilde d_\nu^{({\rm sc})}(E) = \frac
{1}
{{\rm i}\hbar N} \sum_j &&
\frac{T_j^{\rm (p)}}{\kappa_j \sqrt{|\det(M_j-I)|}} \nonumber\\
&&\times\; \chi_\nu(g_j)
\exp\left({\rm i}\frac{S_j}{\hbar} - {\rm i}\mu_j\frac{\pi}{2}\right)
\end{eqnarray}
The sum now runs over generalized period orbits $j$. Their end point is not
necessarily identical with the starting point, but must be mapped to it by
some group element $g_j$. The corresponding term in Eq.~(\ref{symprotrace})
then contains the character $\chi_\nu(g_j)$ as an extra, non-classical phase
factor.  A correction of the amplitude for orbits that coincide with symmetry
lines is effected by $\kappa_j$ \cite{rob}. As usual, $T_j^{\rm (p)}$, $M_j$,
$S_j$, $\mu_j$ denote primitive period, stability matrix, classical action,
and Maslov index, respectively, of orbit $j$.

The r\^ole of the generalized periodic orbits becomes even more transparent in
the analogous trace formula for the group-element-specific density,
\begin{eqnarray}
\label{elspectrace}
d_n^{({\rm sc})}(E) = \frac{1}{{\rm i}\hbar N} \sum_j &&
\frac{T_j^{\rm (p)}}{\kappa_j \sqrt{|\det(M_j-I)|}} \nonumber\\
&&\times\;  \delta(g_j,g_n)
\exp\left({\rm i}\frac{S_j}{\hbar} - {\rm i}\mu_j\frac{\pi}{2}\right).
\end{eqnarray}
The delta function in the second line equals unity if its arguments coincide
and vanishes otherwise. It selects orbits $j$ whose endpoints are connected by
$g_n$. They mediate transport from the original segment to its image, with the
restriction that initial and final points are exactly related by the symmetry.

The interpretation that spectral quantities associated with $g_n$ describe
transport from an original space segment to its image under $g_n$ is borne out
quite explicitly by the form factors. A semiclassical expression for the
$K_n(\tau)$ can be derived by substituting into Eq.~(\ref{formfacdef}) the
trace formula (\ref{elspectrace}), Fourier transformed to the time domain as
in Eq.~(\ref{elspecamp1}). Within the diagonal approximation with respect to
pairs of generalized periodic orbits \cite{ber,dit1,dit2,dit3,arg}, which is
valid for times $t \ll t_{\rm H}$, one obtains,
\begin{equation}
\label{formfacscl}
K_n^{({\rm sc})}(\tau) =
\gamma_n \tau P(g_n,\tau t_{\rm H}),
\qquad \mbox{$\tau \ll 1$.}
\end{equation}

Equation (\ref{formfacscl}) relates the form factors to the classical
probability $P(g_n,t)$ to return in time $t$ to a phase-space point related to
the starting point by $g_n$. The contribution of repetitions of shorter periodic
orbits has been neglected in Eq.~(\ref{formfacscl}). By introducing a global
degeneracy factor $\gamma_n$ to account for antiunitary symmetries like
time-reversal invariance, we ignored the occurrence of self-retracing orbits.
This factor takes the value 2 if orbits that are periodic modulo $g_n$ are
generically time-reversal degenerate, and 1 else. A non-trivial dependence of
$\gamma_n$ on $n$ can occur, e.g., in periodic systems with $N \ge 3$ unit cells
\cite{dit1}. There, time-reversal invariance is generally broken for orbits with
winding numbers $n\,{\rm mod}\,N \neq 0$, $N/2$, due to Bloch phases that are
not real.

In the spirit of the known classical sum rules for ergodic systems \cite{han},
we assume that the generalized periodic orbits are not distinct from the
generic non-periodic ones in their average spreading. We can then relate the
$P(g_n,t)$ to the classical propagator $p({\bm r}',{\bm r};t)$ (the integral
kernel of the Frobenius-Perron operator) by a phase-space integration over the
fundamental domain,
\begin{equation}
\label{cltrace}
P(g_n,t) = \int_{\rm fd} {\rm d}r\, p(g_n({\bm r}),{\bm r};t),
\end{equation}
where ${\bm r} = ({\bm p},{\bm q})$ denotes a phase-space point within the
fundamental domain on the energy shell. In case that the chaotic coverage of the
single cells is homogeneous, the classical propagator depends on $g_n$ but not
on ${\bm r}$ and can thus be expressed by the coarse-grained probabilities
defined in section \ref{sec:cls} \cite{dit1}. In this case the integration in
Eq.~(\ref{cltrace}) becomes trivial and results in
\begin{equation}
\label{clhomotrace}
P(g_n,t) = P_n(t)\,.
\end{equation}
For the group-element-specific form factors defined for two-cell systems,
Eq.~(\ref{twoelformfac}), the semiclassical expression finally reads
\begin{eqnarray}
\label{twoelffscl}
K_{0/1}^{({\rm sc})}(\tau) &=&
\gamma \tau P_{0/1}(\tau t_{\rm H})\\
&=&
\label{symffscl}
\frac{\gamma\tau}{2} (1 \pm {\rm e}^{-2\lambda t_{\rm H} \tau}),
\qquad \mbox{$\tau \ll 1$.}
\end{eqnarray}
We have used here that in symmetric two-cell systems, the degeneracy factor
$\gamma$ does not depend on $g_n$. 

A detailed investigation of the phase-space coverage for a specific double
billiard \cite{kob} shows that---for a finite time depending on the special
properties of the employed model---there can be small deviations from the
homogeneity assumed in the derivation of Eq.~(\ref{twoelffscl}). Similar
restrictions of our theory may arise from the neglect of marginally stable
(bouncing-ball) orbits.  However, all the approximations discussed so far are
standard within a semiclassical theory for two-point correlations, and
although they cannot be rigorously justified, they are sufficient to reproduce
most of the available numerical data \cite{ber,dit1,dit2,dit3,arg}. 

The most important limitation in this respect is due to the diagonal
approximation for systems with time-reversal invariance, $\gamma=2$. It is
well known that this approxi\-mation reproduces only the slope near $\tau=0$
for systems with a single chaotic cell, and it is not surprising that we
observe the same for a two-cell system when we compare (\ref{symffscl}) to
Eq.~(\protect\ref{tc-rmt}). The semiclassical result deviates exactly by the
same factor $\gamma\tau/K_{\rm RMT}(\tau)$ known from simple ergodic systems
\cite{ber}.

In the  remaining paragraphs of this section, we will discuss briefly the
influence of a small breaking of the symmetry of the two billiard cells. It is
clear that the overshoot of $K_0(\tau)$ over its asymptotic value, as described
by Eq.~(\ref{symffscl}), is the result of the symmetry. For an asymmetric double
billiard, this overshoot should vanish altogether, leaving the steeper initial
rise of the form factor implied by Eq.~(\ref{symffscl}) as the only spectral
signature of the restricted exchange between the cells. However, we expect that
there exists a continuous crossover, as a function of some parameter, that
expresses the degree of symmetry breaking.

A semiclassical approach that extends the above arguments to the case of a
weakly broken symmetry has been presented in \cite{dit4}. It is based on the
idea that the contributions of a given set of N symmetry-related periodic orbits
to the form factor will no longer be N-fold degenerate, but can still be
completely included, without resorting to a diagonal approximation within this
set. The following assumptions have to be made to justify this strategy: (i),
the perturbation is sufficiently weak not to destroy the structural stability of
the periodic orbits, i.e., no periodic orbits disappear or are created, as
compared to the unperturbed system; (ii), only the change $\Delta S$ in action
has to be taken into account because it appears in the exponential, while the
changes in amplitude and period can be neglected; (iii), the changes in action
result from many statistically independent perturbations of the orbit so that,
by the central limit theorem, they can be considered as Gaussian random
variables with zero mean and variance
\begin{equation}
\<(\Delta S)^2\> = \delta^2 \tau.
\end{equation}
The constant $\delta$ will serve as the basic parameter for the degree of
symmetry breaking. The proportionality to time reflects the accumulation of
squared action changes over the length of the periodic orbit.

Under these assumptions, the influence of symmetry breaking can be expressed as
an effective degeneracy relating the form factor for broken symmetry to the
corresponding one for the symmetric system (in the present context, we are left
with the ``symmetry-insensitive'' $K_0(\tau)$). For a two-cell system, we find
\begin{equation}
\label{asymffrel}
K_0^{({\rm asym})}(\tau,\delta) =
\frac{1}{2}(1 + {\rm e}^{-\delta^2\tau}) K_0(\tau).
\end{equation}
Inserting $K_0^{({\rm sc})}(\tau)$ from Eq.~(\ref{symffscl}), this implies
\begin{equation}
\label{asymffabs}
K_0^{({\rm asym})}(\tau,\delta) = \frac{1}{2}(1 + {\rm e}^{-\delta^2\tau})
(1 + {\rm e}^{-2\lambda t_{\rm H} \tau}).
\end{equation}
Since this result involves the same semiclassical approximations as we made in
the derivation of $K_0^{({\rm sc})}(\tau)$, its validity is likewise restricted
to the short-time regime $\tau \ll 1$. Within this regime, it provides the
seaked interpolation between the symmetric limit, $\delta^2 \ll 1$, and the
limit of totally broken symmetry, $\delta^2 \appgtr 1$. Due to the assumptions
enumerated above, in particular that of structural stability, we expect
Eq.~(\ref{asymffrel}) to become unreliable also in the latter limit.

\section{Long-time regime}
\label{sec:qum}
In the semiclassical time range, we succeeded to express the form factors for
all parameter regimes, from two-cell systems without significant separation of
the cells to pairs of nearly uncoupled cells, by a single expression,
Eq.~(\ref{symffscl}). We cannot achieve this generality for the regime of long
times $t \appgtr t_{\rm H}$. The case of weakly coupled ($\lambda t_{\rm H}
\ll 1$) symmetric billiards, to be considered here, requires additional input
besides the classical information contained in Eq.~(\ref{symffscl}). At the
same time, this is the most interesting situation because only here,
quasidegenerate doublets occur with a splitting much smaller than their
typical separation.

As a starting point for an alternative approach valid in the long-time regime,
we return to the exact definition of the group-element-specific amplitudes
$a_{0/1}(\tau)$. Specializing Eq.~(\ref{elspecamp1}) to the two-cell case and
inserting the definition (\ref{symredden}) of $d_{0/1}(E)$, we obtain
\begin{equation}
\label{twoelamp}
a_n(\tau) = \frac{1}{2} \sum_{\nu=0}^1 {\rm e}^{\pi{\rm i}n\nu}
\sum_{\alpha = 1}^{N_{\rm d}} {\rm e}^{-2\pi{\rm i}\tau r_{\alpha,\nu}},\quad
n = 0,1.
\end{equation}
Here, $N_{\rm d}$ is the number of doublets in the spectrum, i.e., half the
total number of levels. We have used the fact that the inverse characters for
the twofold reflection or translation group can be concisely written as
$(\chi^{-1})_{\nu,n} = {\rm e}^{\pi{\rm i}n\nu}$, with $\nu = 0$, 1 corresponding to
the symmetric and the antisymmetric representations, respectively. In these
representations, returning to the symbols `$+$' and `$-$', $r_{\alpha,0/1} =
r_{\alpha,\pm} = \< d_{\rm fd}\> E_{\alpha,\pm}$ denote the scaled
eigenenergies.

We introduce the concept of doublets by writing the eigenenergies as
\begin{equation}
\label{doublets}
r_{\alpha,\pm} = R_{\alpha} \pm r_{\alpha}\,.
\end{equation}
The long-time limit of the form factors for the canonical random-matrix
ensembles is usually derived under the assumption that the full phases $\tau
r$ are random for $\tau \gg 1$. Likewise, we here assume the analogous phases
$\tau R_\alpha$ contributed by the doublet midpoints to be random in the
long-time limit. Upon squaring the amplitudes $\tilde a_\pm(\tau)$ to obtain
the corresponding form factors, this amounts to a diagonal approximation with
respect to the index $\alpha$,
\begin{equation}
\label{fftunsplit}
K_{0/1}(\tau) = \frac{1}{2} \pm \frac{1}{2N_{\rm d}}
\sum_{\alpha = 1}^{N_{\rm d}} \cos(4\pi\tau r_{\alpha}).
\end{equation}
In fact, Eq.~(\ref{fftunsplit}) can be derived also if the two-cell system is
unrolled to an infinite chain and the doublets are considered as points of
continuous bands, see \ref{app:ban}.

If $N_{\rm d}$ is sufficiently large, we can replace the sum over $\alpha$ by
an integral and consider the integration as a Fourier transformation, to
obtain
\begin{equation}
\label{fftundis}
K_{0/1}(\tau) =
\frac{1}{2} \left(1 \pm p_{\rm d}(2\tau)\right),
\end{equation}
with
\begin{equation}
\label{tundisft}
p_{\rm d}(\tau) = \int_0^{\infty}{\rm d}r\,
\cos(2\pi r\tau) p_{\rm d}(r).
\end{equation}
This is the Fourier transform of the distribution of doublet splittings
\begin{equation}
\label{splitdist}
p_{\rm d}(r) = \sum_{\alpha = 1}^{N_{\rm d}}
\delta (r - |r_{\alpha,-} - r_{\alpha,+}|)\,.
\end{equation}
Forming the difference of the two form factors,
\begin{equation}
\label{ffdiff}
p_{\rm d}(\tau) = K_0(\tau/2) - K_1(\tau/2),
\end{equation}
we see that this equals the time-domain splitting distribution in its
long-time, or equivalently, low-energy limit. Given that the form factors
contain information merely on two-point correlations irrespective of symmetry,
it is actually surprising that they can be related, as in Eq.~(\ref{ffdiff}),
to a quantity that requires an unambiguous identification of doublets. This
can be explained by the fact that we had to assume in the derivation that the
midpoints $R_\alpha$ of the doublets are statistically independent of their
splittings $r_\alpha$, which requires a clear separation of scales between
splittings and spacings of doublets. Indeed, Eq.~(\ref{ffdiff}) ceases to be
valid for $\tau \applss 1$, corresponding to the regime of large 
$r_\alpha\stackrel{>}{\sim}1$.

We do not have any semiclassical access to $p_{\rm d}(r)$. In order to
nevertheless make some progress, we shall resort to results of random-matrix
theory on the distribution of resonance widths, and argue that the doublet
splittings obey a similar distribution.

\begin{figure}
\centerline{\psfig{figure=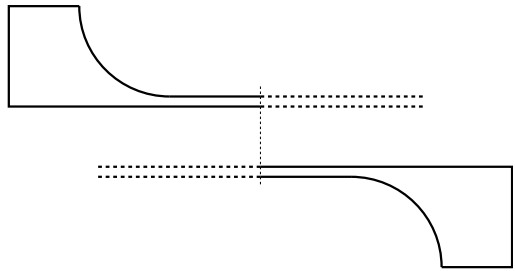,width=7cm}}
\caption{\label{hdb-fig} Two identical scattering systems are obtained when
the channel connecting the two halves of a symmetric two-cell billiard is
replaced by a semi-infinite waveguide of constant width. We argue that small
doublet splittings between states of the two-cell billiard correspond to
narrow resonances of the corresponding pair of scattering systems (see text).
}
\end{figure}
Suppose the channel between the two cells to be replaced by a semi-infinite
duct of constant width, so that two single cells remain, each with a small
opening that couples its interior to the continuum in the duct. This situation
is illustrated with a symmetric two-cell billiard in Fig.~\ref{hdb-fig}.  The
spectra of the open single cells will then exhibit narrow resonances at
roughly the same energies where the corresponding closed two-cell system shows
doublets. It is plausible that the doublet splittings of the two-cell
configuration are related to the resonance widths in the single-cell setup, at
least in a statistical sense. Indeed, this assumption is often made, e.g., in
nuclear theory, and supported by semiclassical and random-matrix arguments. In
short, it is justified by the fact that both quantities, widths and
splittings, can be expressed by the same wavefunction overlaps and therefore
should obey the same distribution.

Here we are interested in the case of very few open channels in the connecting
section, since this is required for the formation of narrow doublets. In this
limit, we cannot expect random-matrix theory to remain valid. However, taking
the point of view of the scattering approach to quantization \cite{dor}, we
show in the following paragraphs that there is still a close relation, though
not an exact identity, between the respective distributions of widths and
splittings.

We shall consider specifically the wavenumber regime where there is just a
single open channel in the connecting section. This case is realized in the
majority of the numerical examples below. Here, the calculation is
particularly straightforward and transparent, because the scattering in the
cell is described by a $1\times 1$ scattering ``matrix'' (since there is only
one opening, all incoming waves leave as reflected waves). As we are
interested in isolated narrow resonances, we assume their widths to be small
as compared to their separations. The $S$ matrix can then be written in the
form
\begin{equation}
\label{onedimsmatrix}
S(r) = {\rm e}^{{\rm i}\theta} \frac{r-r_{0}-{\rm i}g}{r-r_{0}+{\rm i}g}\, ,
\end{equation}
where $r_0-\i g$ denotes the position of the resonance pole in the complex
energy plane, again in units of the mean level spacing.
The quantization condition in terms of $S(r)$ is \cite{dor}
\begin{equation}
S(r_{\pm}) = \pm 1,
\end{equation}
the upper sign referring to the case of symmetric states (subscript `$+$'),
the lower one to antisymmetric states (`$-$'). Inserting
Eq.~(\ref{onedimsmatrix}) gives the eigenvalues
\begin{eqnarray}
r_+(\theta) &=& r_{0}+g \cot \theta/2,\\
r_-(\theta) &=& r_{0}-g \tan \theta/2,
\end{eqnarray}
for the symmetric and antisymmetric cases, respectively, separated by the
splitting
\begin{equation}\label{r_theta}
r(\theta) = |r_-(\theta) - r_+(\theta)| = \frac{2g}{|\sin \theta|}\,.
\end{equation}
This function is $\pi$ periodic. In the interval $0 \leq \theta < \pi$, it has
a minimum at $\theta = \pi/2$, with functional value $r(\pi/2) = 2g$. It
diverges at $\theta = 0$, $\pi$.

This analysis already exhibits the essential facts to be demonstrated: There
is a connection between resonance widths and doublet splittings, but it
depends on the unknown value of the total phase $\theta$ of the S matrix. As
no value of $\theta$ is singled out a priori, we assume equidistribution of
the total phase. Under this condition, the main contribution to the
distribution of the splittings comes from $r_{\rm s} \appgtr 2g$. This is the
simple relation between doublet splittings and resonance widths we
seek. Quantitatively, we find the probability density
\begin{equation}
\label{condspdist}
p_{\rm d}(r|g) = \frac{4g}{\pi r^2}
\left(1 - \left(\frac{2g}{r}\right)^2\right)^{-1/2}\qquad
(r\ge 2g)\,.
\end{equation}
for the splittings. It is normalized to unity, but already its first moment
diverges. Indeed, as we started from the assumption of small splittings, we
cannot expect the result to be valid for large splittings. The missing cutoff
will be given by our semiclassical considerations which cover the
complementary regime of large splittings.

The distributions of wave-function amplitudes and of resonance widths are
among the established principal results of random-matrix theory \cite{bro}.
Even if details of their application to quantum chaotic scattering are still
under study, we can, for the present purposes, adopt the canonical
random-matrix results for $p(g)$, the probability density of the resonance
widths, and substitute them to obtain the unconditional splitting distribution
$p_{\rm d}(r)$.

Its general relation with the conditional $p_{\rm d}(r|g)$ and $p(g)$ reads
\begin{equation}
p_{\rm d}(r) = \int_0^{\infty} {\rm d}g\, p(g) p_{\rm d}(r|g).
\end{equation}
In order to get back from the energy to the time domain, we perform a Fourier
transformation of $p_{\rm d}(r)$, cf.\ Eq.~(\ref{tundisft}). Inserting the
explicit expression (\ref{condspdist}) for $p_{\rm d}(r|g)$, we obtain
\begin{equation}\label{p_d_tau}
p_{\rm d}(\tau) =
\frac{1}{\pi} \int_0^{1} {\rm d}x\,\frac{2}{\sqrt{1 - x^2}}\,
p\left(\frac{2\tau}{x}\right),
\end{equation}
where $p(\tau) = \int_0^{\infty} {\rm d}g \cos(2\pi g\tau) p(g)$, in
turn, is the Fourier transform of the distribution of resonance widths.

For time-reversal-invariant systems, the resonance widths obey a Porter-Thomas
distribution \cite{bro}
\begin{equation}
\label{ptdist}
p_{\rm PT}(g) = \frac{{\rm e}^{-g/2\< g\>}}
{\sqrt{2\pi\< g\> g}}\,.
\end{equation}
The integral obtained by inserting the Fourier transform 
\begin{eqnarray}\label{p_pt_tau}
p_{\rm PT}(\tau)&=&\sqrt{1+\sqrt{1+[4\pi\<g\>\tau]^2}\over 2(1+[4\pi\<g\>\tau]^2)}
\end{eqnarray}
into (\ref{p_d_tau}) can be used for a numerical computation of the form
factor. In the long-time limit we find from an asymptotic expansion of 
(\ref{p_pt_tau})
\begin{equation}\label{asym-trs}
p_{\rm d}(\tau)
={1\/4\pi}{\Gamma(3/4)\/\Gamma(5/4)}{1\/\sqrt{\<g\>\tau}}+
{\rm O}\(\(\<g\>\tau\)^{-3/2}\)\,.
\end{equation}
If time-reversal invariance is broken, the resonance widths are exponentially
distributed,
\begin{equation}
\label{exdist}
p_{\rm exp}(g) = \frac{1}{\< g\>} {\rm e}^{-g/\< g\>},
\end{equation}
\begin{equation}\label{p_exp_tau}
p_{\rm exp}(\tau)={1\/1+\(2\pi\<g\>\tau\)^2}
\end{equation}
and accordingly
\begin{equation}
\label{uncondexft}
p_{\rm d}(\tau) = \frac{1}{1 +
4\pi\< g\>\tau \sqrt{1 + (4\pi\< g\>\tau)^2} +
(4\pi\< g\>\tau)^2}.
\end{equation}
For $\< g\>\tau \gg 1$ we find $p_{\rm d}(\tau)\to (4\pi\<g\>\tau)^{-2}/2$.

Note that for both, exponential and Porter-Thomas distribution, the asymptotic
behaviour of the doublet splittings for large time/small energy is equivalent to
the corresponding resonance-width distribution up to a constant prefactor which
relates the mean doublet splitting to $\<g\>$. This constant---it equals
$2\pi\Gamma^2({5\/4})/\Gamma^2({3\/4})\approx 3.44$ with and $\sqrt{8}\approx
2.82$ without time-reversal invariance---is somewhat above 2 as anticipated from
Eq.~(\ref{r_theta}).

\begin{figure}
\centerline{\psfig{figure=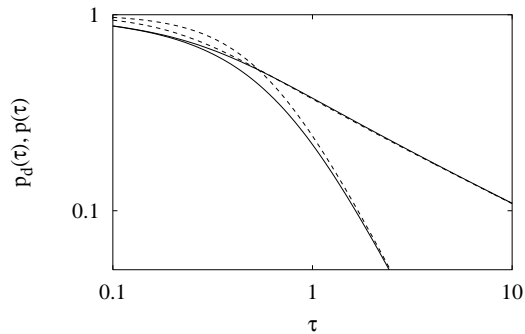,width=7cm}}
\caption{\label{splitres} Time-domain splitting distribution
Eq.~(\protect\ref{p_d_tau}) for symmetric two-cell systems with a single open
channel in the connector, in the presence of time-reversal invariance (upper
solid line) and in its absence (lower solid line). For comparison, the broken
lines show the Fourier-transforms of the corresponding resonance
distributions, i.e., of the Porter-Thomas distribution
Eq.~(\protect\ref{p_pt_tau}), above, and of the exponential distribution
Eq.~(\protect\ref{p_exp_tau}), below. The parameter value common to all curves
is $\<g\> = 0.1$.}
\end{figure}
In Fig.~\ref{splitres}, we compare the Fourier-transformed unconditional
splitting distribution Eq.~(\ref{p_d_tau}) in the presence and absence 
of time-reversal invariance to the corresponding distribution of resonance
widths for $\<g\> = 0.1$. We see that for $\< g\>\tau \appgtr 1$, the
deviation between the two is not dramatic and conclude that the resonance
distribution $p(g)$, Porter-Thomas or exponential, is the crucial input for
$p_{\rm d}(\tau)$, while it is quite robust against changes and approximations
entering via $p_{\rm d}(r|g)$.

Now we return to our main line of reasoning and attempt a matching of the
short-time (large-separation) with the long-time (small-splitting) regime of
the form factors. This will simultaneously allow us to calibrate the as yet
undetermined parameter $\< g\>$ of the resonance distribution with respect to
the classical decay rate $\lambda$. We shall present this calculation only for
the simpler case of broken time-reversal invariance. If time-reversal symmetry
is obeyed, the bad performance of the diagonal approximation at the Heisenberg
time makes an analogous procedure more problematic. We will discuss this in
connection with our numerical results in Sections \ref{ssec:sinai2} and
\ref{ssec:qgr}.

From the semiclassical side, Eq.~(\ref{symffscl}), we find at the matching
point $\tau = 1$,
\begin{equation}
\label{ffshorttimematch}
K_{0/1}(1) = \frac{1}{2}
\left(1 \pm {\rm e}^{-2\lambda t_{\rm H}}\right),
\end{equation}
while from the long-time side, substituting Eq.~(\ref{uncondexft}) in
Eq.~(\ref{fftundis}), we have
\begin{equation}
\label{fflongtimematch}
K_{0/1}(1) = \frac{1}{2}
\left(1 \pm \frac{1}{1 + c\sqrt{1 + c^2} + c^2}\right),
\end{equation}
introducing the shorthand $c = 8\pi\< g\>$. These equations are
consistent with one another if
\begin{equation}
{\rm e}^{-2\lambda t_{\rm H}} = \frac{1}{1 + c\sqrt{1 + c^2} + c^2},
\end{equation}
or, resolving for $c$,
\begin{equation}
\label{relsplitdecay}
c = \frac{1 - {\rm e}^{-2\lambda t_{\rm H}}}
{\sqrt{{\rm e}^{-2\lambda t_{\rm H}}(2 - {\rm e}^{-2\lambda t_{\rm H}})}}.
\end{equation}
For $\lambda t_{\rm H}\ll 1$ (narrow connecting channel or weak coupling),
proportionality $\<g\>\approx \lambda t_{\rm H}/4\pi$ results. It represents a
simple relation between the parameter of the quantum-mechanical splitting
distribution and the classical time scale of equilibration between the cells.
\begin{figure}
\centerline{
\psfig{figure=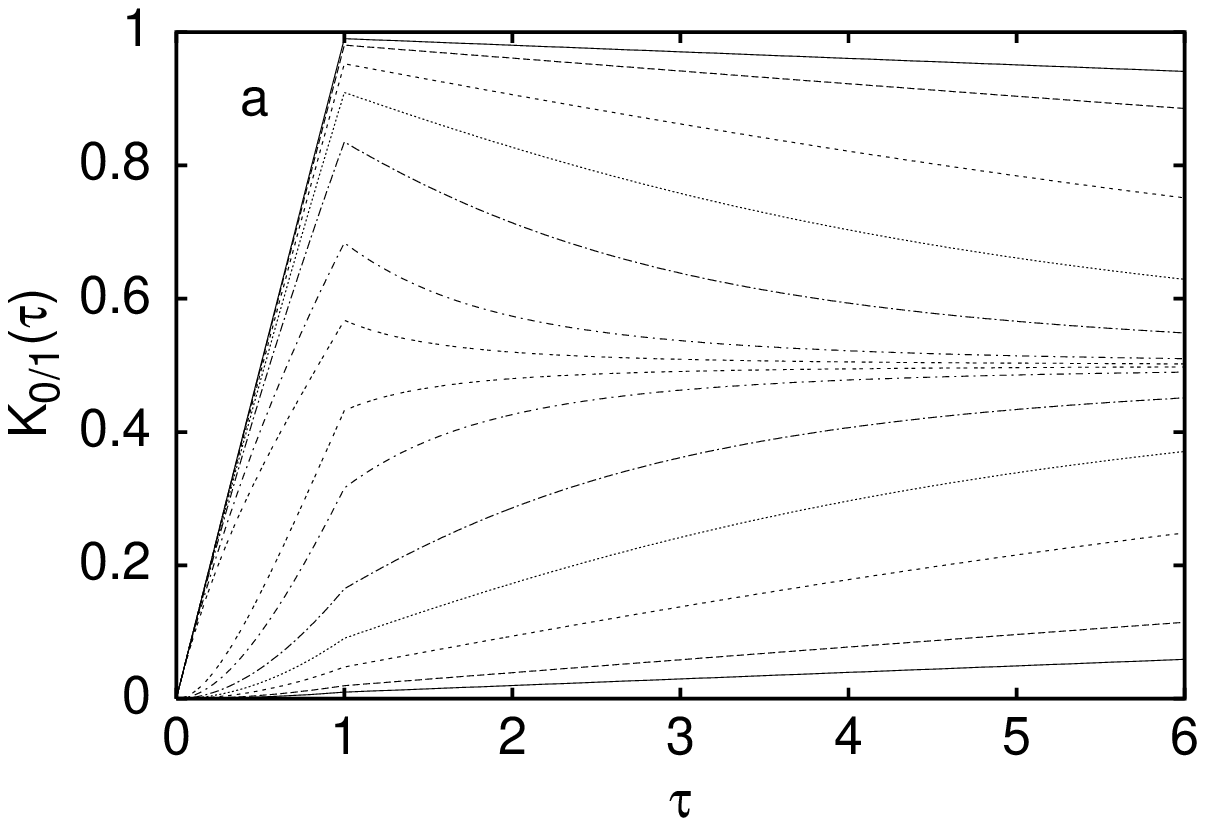,width=6cm}
\psfig{figure=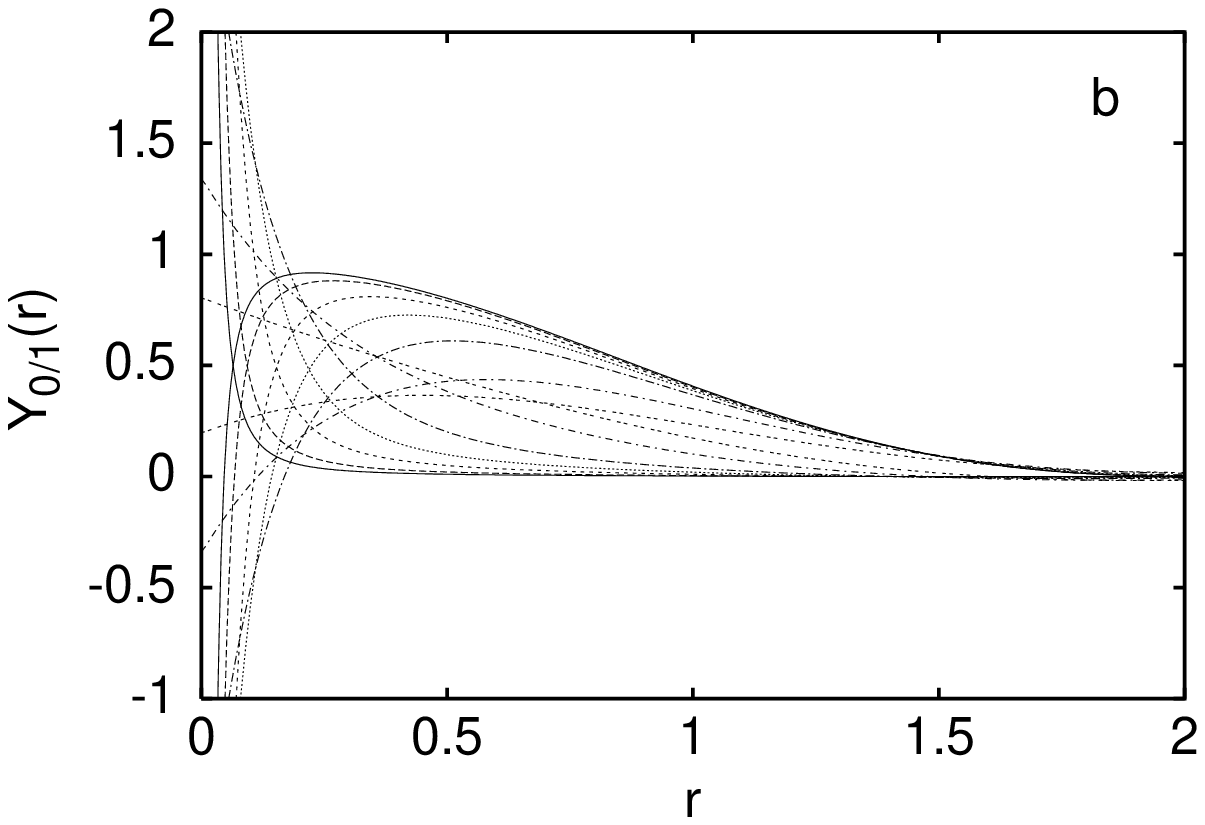,width=6cm}
}
\caption{\label{ffex} 
Group-element-specific form factors (a), as described by
Eqs.~(\protect\ref{symffscl}) and (\protect\ref{ffexlongtime}), and
corresponding cluster functions (b), for the case of broken time-reversal
invariance. In (a), the upper curves show $K_0(\tau)$, the lower curves
$K_1(\tau)$. In (b), the graphs with positive initial slope correspond to
$Y_0(r)$, those with negative initial slope to $Y_1(r)$.  From the outmost to
the innermost pair of curves, the decay rate takes the values $\lambda t_{\rm H} =
0.01$, 0.02, 0.05, 0.1, 0.2, 0.5, 1.0. Graphs of form factors and cluster
functions for equal values of $\lambda t_{\rm H}$ share a common line signature.}
\end{figure}
  
We state the full time dependence of the form factors in the long-time regime,
again using the abbreviation $c$ for the sake of conciseness,
\begin{equation}
\label{ffexlongtime}
K_{0/1}(\tau) = \frac{1}{2} \left(1 \pm
\frac{1}{1 + c\tau\sqrt{1 + c^2\tau^2} + c^2\tau^2}\right).
\end{equation}
In Fig.~\ref{ffex}, we give a synopsis of $K_{0/1}(\tau)$ (a) and $Y_{0/1}(r)
= \int_0^\infty{\rm d}\tau\,[1 - K_{0/1}(\tau)]\cos(2\pi r\tau)$ (b), for
values of the decay constant ranging from $\lambda \ll 1$ to $\lambda \approx
1$. The figure illustrates the crossover of the spectral two-point
correlations from the regime of almost immediate equidistribution between the
cells, $\lambda \appgtr 1$, where the two-point statistics barely deviates
from the corresponding GOE or GUE prediction (for the figure, we have chosen
the case of broken time-reversal invariance where the semiclassical
approximation to the random-matrix form factor is exact), to the regime of
weak coupling, $\lambda \ll 1$, with $K_0(\tau)$ rising to a marked peak near
$\tau = 1$.

\section{Models and numerical results}
\label{sec:num}
In the following subsections, we shall introduce five quite diverse models that
allow to construct systems with two coupled compartments. The numerical results
obtained for these models serve to illustrate and check various aspects of the
theory developed above.

\subsection{The Z-shaped billiard}
\label{ssec:sinai1}
\label{sssec:cl}
\begin{figure}[htb]
\centerline{\psfig{figure=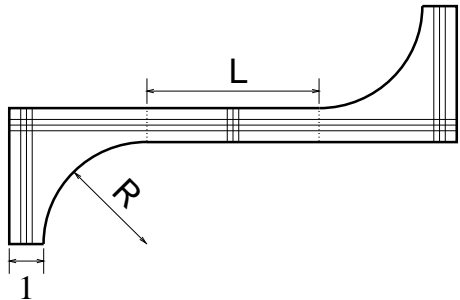,width=6cm}}
\caption{\label{db-fig} 
The Z-shaped billiard and the three different families of bouncing-ball orbits.}
\end{figure}
\begin{figure}[htb]
\centerline{\psfig{figure=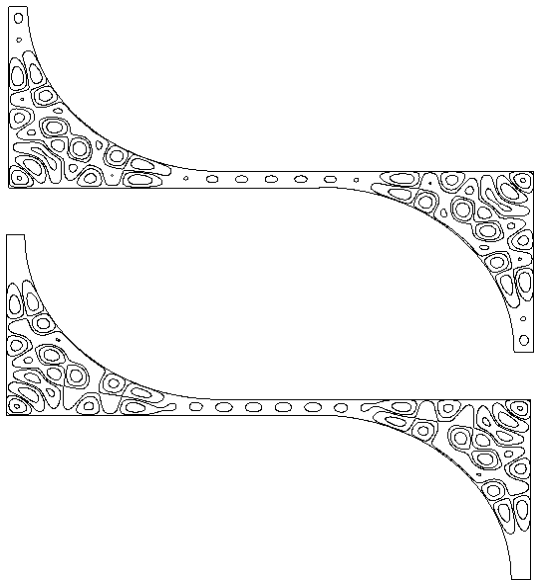,width=7cm}}
\caption{\label{cont-fig} 
Contour plot of the absolute square of a pair of eigenfunctions of the double
billiard with quasidegenerate energies. The state with even symmetry has
$k=3.7569$ (top), the antisymmetric state has $k=3.7576$ (bottom). The
geometric parameter values are $R = 10$ and $L = 5$.}
\end{figure}

We construct a two-cell billiard from two quarters of a Sinai
billiard \cite{sin} and a straight channel such that the resulting shape
resembles the letter Z \cite{kob} (Fig.~\ref{db-fig}). The width of this channel will
serve as the basic length unit. The remaining parameters of the billiard are
then the length $L$ of the channel and the common radius $R$ of the
quarter-circle sections of the boundary.

Since the billiard boundary consists exclusively of defocussing and neutral
components, the classical dynamics is ergodic and mixing \cite{sin}. Hence
we can assume that Eq.~(\ref{mastersol}) holds to a good approximation,
although for finite time systematic deviations from ergodicity, e.g., due to
the presence of bouncing-ball orbits (Fig.~\ref{db-fig}), can be observed. In
the following, we neglect such effects which have been studied in detail in
\cite{kob}.

The employed quantization scheme is described in \ref{sssec:qm}.
Figure \ref{cont-fig} shows a representative example of a pair of
eigenfunctions of the double billiard with quasidegenerate eigenenergies.

\begin{figure}
\centerline{\psfig{figure=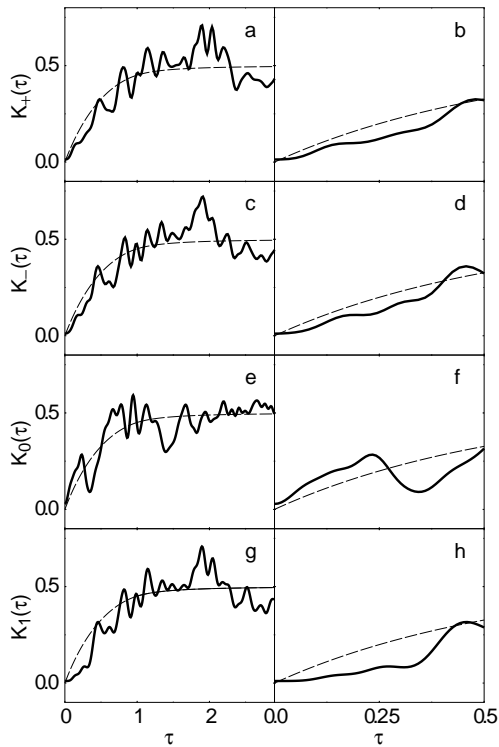,width=7cm}}
\caption{\label{ff-fig} Time evolution of various form factors (heavy lines),
compared to random-matrix theory (broken). The upper two rows show the
symmetry-projected form factors (panels a -- d), the lower ones the
group-element-specific form factors (e -- h). The right column consists of
blowups of the short-time regime of the data shown on the left.  The geometric
parameter values are $R = 5$ and $\<L\> = 6.7$.}
\end{figure}

Figure \ref{ff-fig} shows a comparison of the various form factors, defined in
Section \ref{sec:fof}, with random-matrix theory. In Fig.~\ref{ff-fig}, we
show the symmetry-projected form factors $\widetilde K_+(\tau)$ (panels a,b)
and $\widetilde K_-(\tau)$ (c,d) (cf.\ Eq.~(\ref{saformfac})), as well as the
group-element-specific ones $K_0(\tau)$ (panels e,f) and $K_1(\tau)$ (g,h)
(cf.\ Eq.~(\ref{twoelformfac})), together with the corresponding predictions
of random-matrix theory.

The right-hand column contains blowups of the short-time regime of the data
shown on the left. The $\widetilde K_{\pm}(\tau)$ follow closely the GOE
prediction for a single cell of the double billiard, as expected. For short
times, $K_0(\tau)$ and $K_1(\tau)$ deviate significantly from the GOE shape,
in the way predicted by semiclassical considerations. The initial slope of
$K_0(\tau)$ is increased by a factor 2, while that of $K_1(\tau)$ vanishes.
However, the amount of data obtained for this model is too small to allow for a
quantitative comparison with the semiclassical theory beyond the vicinity of
$\tau=0$. Moreover our numerical quantization procedure did not allow to
go to parameter values where $\lambda t_{\rm H} \applss 0.5$ such that
$K_0(\tau)$ is expected to overshoot near $\tau = 1$. For the parameter values
underlying the data shown, we have $\lambda t_{\rm H} \approx 4$, so that
classical equilibration between the cells occurs around $\tau \approx 0.125$,
cf.\ Eq.~(\ref{symffscl}).

\subsection{The Sinai billiard}\label{ssec:sinai2}

\begin{figure}
\centerline{\psfig{figure=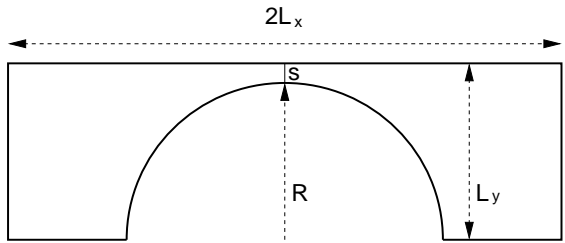,width=8cm}}
\caption{\label{sb:geometry}
One half of the Sinai billiard consisting of a rectangle and an inscribed
semicircle which divides the system into a reflection-symmetric pair of cells. The
numerical data presented in this section correspond to $L_x = 2$, $L_y = 1$,
and $s = 0.05$.}
\end{figure}

Another example of a two-cell billiard is provided by one half of the Sinai
billiard as shown in Fig.~\ref{sb:geometry}. A semicircle of radius $R$
divides a rectangle with side lengths $L_x$, $L_y$, into two parts connected
by an opening of size $s = L_y - R$ along the symmetry axis. According to
Eq.~(\ref{rate}), the width of this constriction determines the classical rate
of transitions between the two cells in the ergodic regime. The geometry of
this billiard differs from that of the system discussed in the previous
subsection in three respects: The single cells are no longer symmetric in
themselves, there is no extra connecting channel of variable length, and the
full configuration has reflection rather than inversion symmetry. The main
advantage is, however, that there exists a more efficient quantization
algorithm \cite{sch}, again based on the scattering approach \cite{dor}. The
reflection symmetry allows to compute the eigenvalues in the two parity
classes separately by requiring Neumann or Dirichlet boundary conditions along
the symmetry axis. We unfold both spectra using the area and circumference
contributions to the mean spectral density of one cell \cite{bal} and arrive
at the scaled energy eigenvalues.

Our theory, developed in Sections \ref{sec:cls} to \ref{sec:qum}, is based on
spectral two-point correlations that indiscriminately include {\em all} level
pairs in the spectrum. The symmetry-based quantization procedure used for the
present model, by contrast, gives us immediate access to the scaled
eigenvalues $r_{\alpha,\pm}$, presorted according to parity. We take this
opportunity to make a few remarks concerning the ``genuine'' doublets, i.e.,
level pairs with identical quantum number $\alpha$ but opposite symmetry, and
their splittings $r_\alpha = r_{\alpha,-} - r_{\alpha,+}$. We emphasize again
that only in the regime of small splittings, statistically independent of the
positions of the doublet centers, the two-point statistics embodied in the
form factors coincides with the distribution of the genuine doublet
splittings, cf.\ Eq.~(\ref{splitdist}). Outside this regime, the two-point
statistics includes separations that are possibly very small but belong to
states labeled by different quantum numbers, and therefore do not contribute
to the splitting distribution. In effect, the two-point statistics is less
restrictive and shows more weight at small separations than the splitting
distribution.

\begin{figure}
\centerline{\psfig{figure=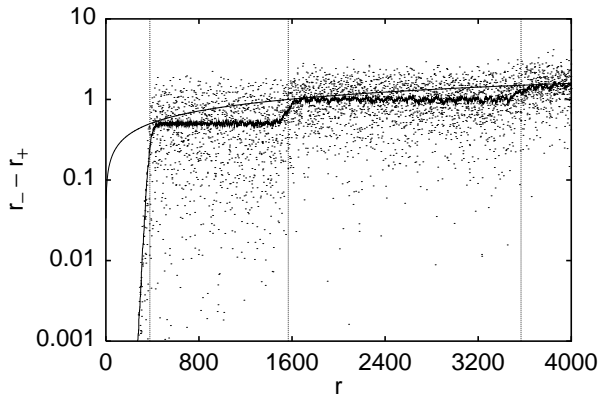,width=8cm}}
\caption{\label{sb:ts} 
Spacings between pairs of unfolded energy eigenvalues with equal quantum
number with respect to a single cell in Fig.~\protect\ref{sb:geometry} and
Dirichlet/Neumann boundary conditions along the symmetry axis. The vertical
lines mark the threshold energies where the first three quantum channels
between the cells open. A running average of the spacings (wiggly line) is
compared to the high-energy approximation (smooth curve).
}
\end{figure}
Figure \ref{sb:ts} shows the individual doublet splittings (dots) and a
running average (wiggly line) as a function of the energy. We observe that the
average splitting essentially depends on the number $\Lambda = [ks/\pi]$ of
open quantum channels in the constriction. For low energy, quasidegenerate
doublets prevail. In particular, below the threshold energy of the first
quantum channel, we have $|\widetilde r_{\alpha}|\ll 1$ for all pairs of
eigenvalues. Because of the analogy with actual tunnel splittings \cite{cre},
we presume that a semiclassical description of the spectrum in this regime
should include also orbits with complex action including the diffractive orbits
studied in \cite{pri}. This question will be investigated elsewhere.

As the energy approaches the opening of the first channel, the mean doublet
splitting increases exponentially, and doublet splittings larger than the mean
level spacing accumulate. Beyond the opening of the second quantum channel,
even the average splitting exceeds the mean spacing. Consequently, for high
energy, the notion of doublets becomes irrelevant for the spectral statistics
of the composite system. It is, though, well suited in the regime of, e.g., a
single open quantum channel, as we will show below.

An approximation to the mean value of the doublet spacing is obtained from the
asymptotic expansion of the mean spectral staircases $\bar N_\pm$ of the two
subspectra. While the leading contribution depending on the area $A$ is the
same for both spectra, the second term depends on the circumference $u$ and
the boundary conditions. For $\hbar=2m=1$, we have
\begin{equation}\label{sb:weyl}
\bar N_-(E) = {A\/4\pi}E-{u\/4\pi}E^{1/2}\qquad
\bar N_+(E) = \bar N_-(E)-{s\/2\pi}E^{1/2}\,.
\end{equation}
With the approximate quantization condition for scaled energy, $\bar
N_\pm(E_{\alpha,\pm}) = \alpha + 1/2$ \cite{aur}, this leads to
\begin{equation}\label{app_ds}
|r_{\alpha}| \approx s \sqrt{r/A\pi}\,,
\end{equation}
which is represented by a smooth solid line in Fig.~\ref{sb:ts}. We see that for
low energy, the approximation (\ref{app_ds}) is correct only in the vicinity of
the channel openings, while the mean splitting is approximately constant between
the thresholds. Accordingly, taking the value predicted by Eq.~(\ref{app_ds}) at
the opening of the first channel $r = A\pi/4s^2$ for the entire subsequent
interval till the next threshold, we find that the mean dimensionless doublet
splitting for one open channel is $1/2$, independently of the size of the hole.
Thus it is already of the order of the mean level spacing of the composite
two-cell system.

It is an important point that this does not restrict the applicability of our
theory: The high probability of large doublet splittings corresponds to the
fact that $p_{\rm d}(r)$, as obtained from Eq.~(\ref{condspdist}), has a diverging
first moment. Nevertheless, its Fourier transform is well behaved.
Beyond the Heisenberg time, where we make use of it, it is essentially
determined by the behaviour of the distribution at small spacings. Indeed,
Fig.~\ref{sb:ts} shows a large number of doublets with a width well below the
mean level spacing, which justifies our approach.

\begin{figure}
\centerline{
\psfig{figure=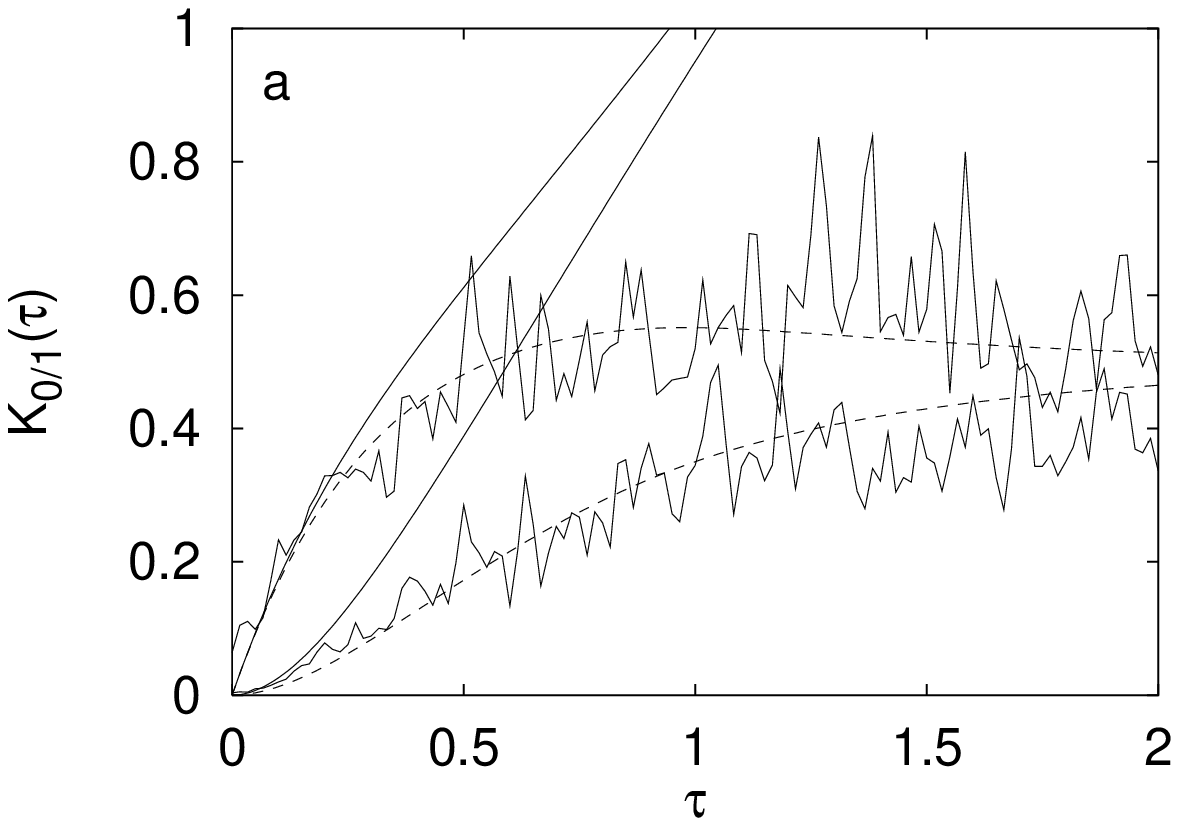,width=6cm}
\psfig{figure=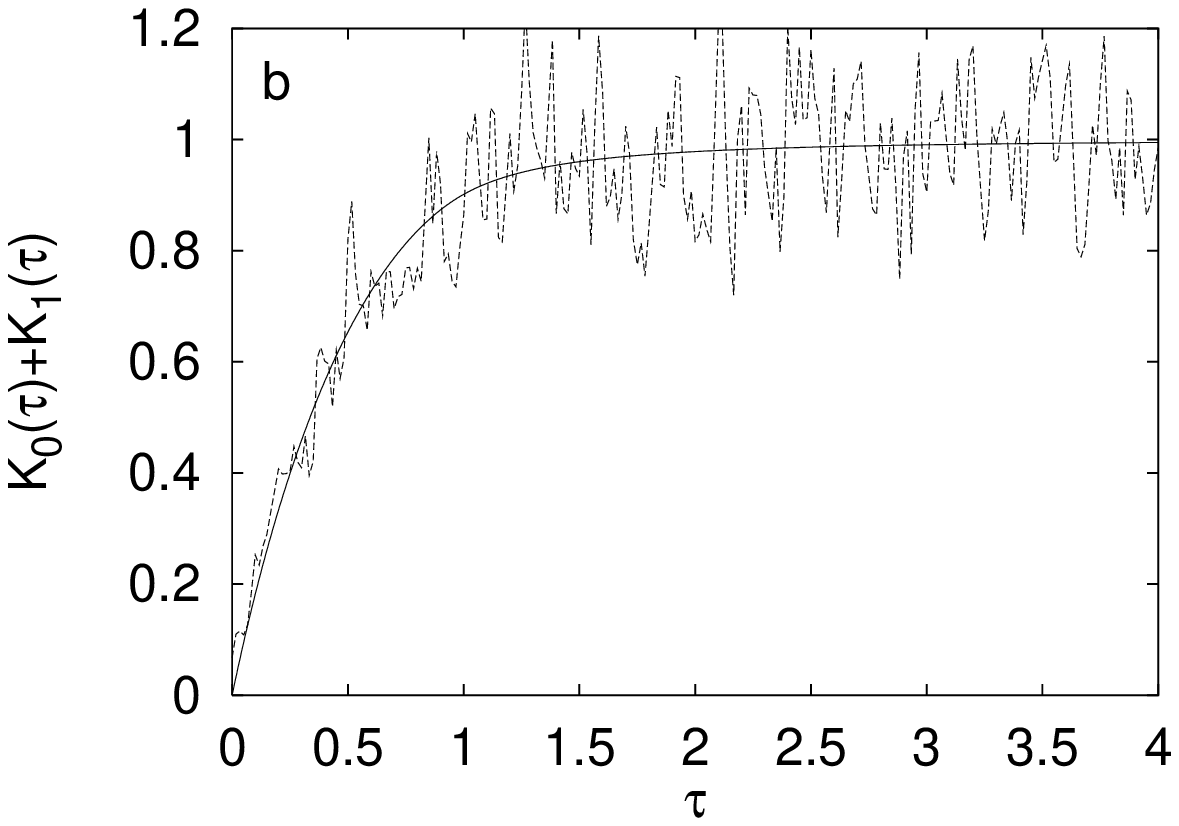,width=6cm}
}
\caption{\label{sb:scl} (a) Form factors $K_{0/1}(\tau)$ for the two-cell
Sinai billiard of Fig.~\protect\ref{sb:geometry}. The smooth solid curves
represent the semiclassical result in diagonal approximation
(\protect\ref{symffscl}), the dashed lines show a fit to the ansatz
(\protect\ref{facff}) which explicitly obeys Eq.~(\protect\ref{tc-rmt}). The
validity of the sum rule (\protect\ref{tc-rmt}) is demonstrated in (b), where
the solid line represents the GOE form factor.
}
\end{figure}
\begin{figure}
\centerline{
\psfig{figure=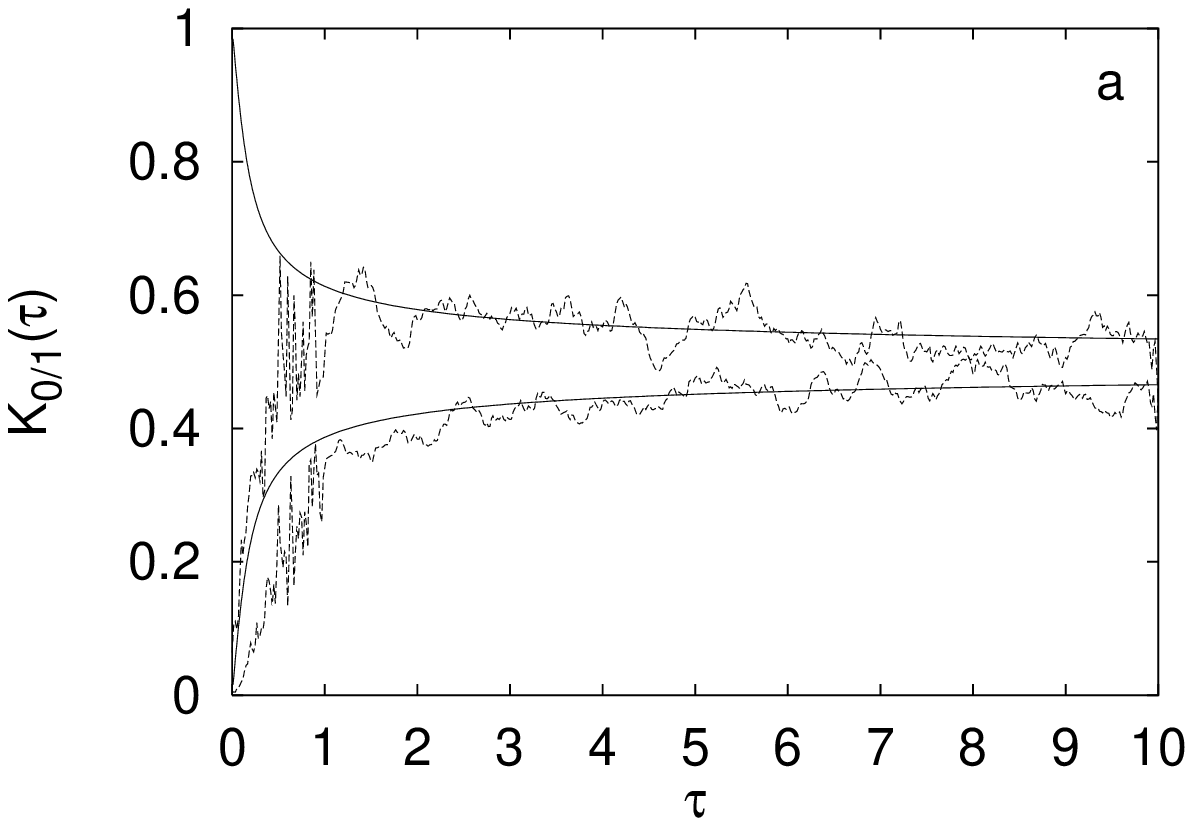,width=6cm}
\psfig{figure=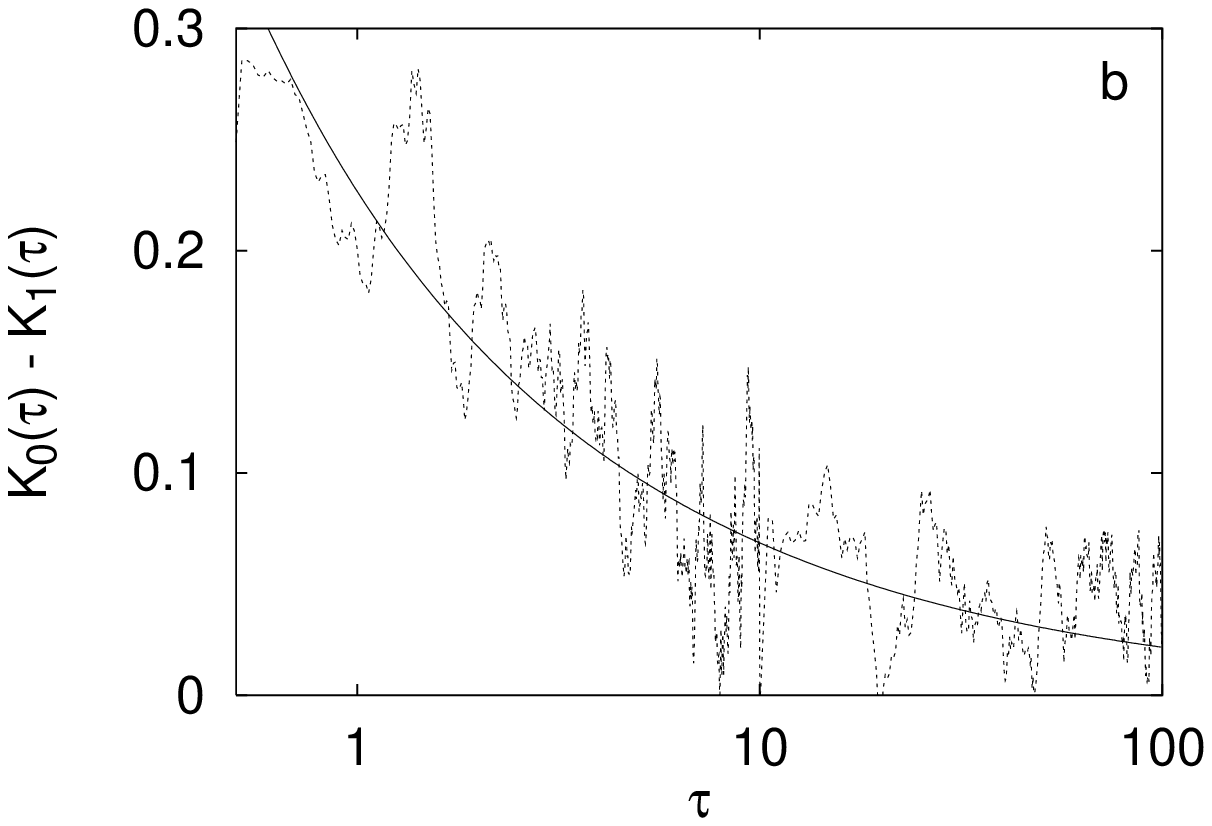,width=6cm}
} 
\caption{\label{sb:long} Long-time behaviour of the group-\-element-specific
form factors for the two-cell Sinai billiard. (a) shows $K_{0/1}(\tau)$
together with the prediction according to Eq.~(\protect\ref{fftundis}).  The
parameter $\<g\>$ was obtained by a fit to the short-time behavior in the
vicinity of the Heisenberg time $\tau=1$ (see text). (b) shows the difference
of the two form factors, which contains all essential information.
}
\end{figure}
In Fig.~\ref{sb:scl}a, we present the form factors $K_{0/1}(\tau)$ obtained
from the 1,187 doublets with one open quantum channel. They were computed
after splitting the spectrum into small intervals of 30 doublets each. For the
parameter $\lambda t_{\rm H}$ entering the semiclassical theory we use the
value obtained from Eq.~(\ref{rate}) with $k$ as at the center of the
considered interval. The dashed line shows the prediction of the semiclassical
diagonal approximation (\ref{symffscl}), which correctly describes the
behaviour of the form factors for small $\tau$, but fails close to the
Heisenberg time $\tau=1$ as discussed at the end of Section \ref{sec:scl} in
connection with the sum rule (\ref{tc-rmt}). The validity of this relation is
demonstrated in Fig.~\ref{sb:scl}b. It is interesting to note that all our
data for systems with time-reversal invariance can be fitted very accurately
(with $\lambda$ as a free parameter) using an ansatz which combines
(\protect\ref{tc-rmt}) and the semiclassical result (\ref{symffscl}) into
\begin{equation}\label{facff}
K_{0/1}(\tau) = K_{\rm RMT}(\tau) P_{0/1}(\tau t_{\rm H})\,.
\end{equation}
We cannot further substantiate this expression analytically. 

Due to the poor outcome of the diagonal approximation in the present case, we
cannot directly determine the mean resonance width from the decay constant by
matching (\ref{symffscl}) and (\protect\ref{fftundis}) at $\tau=1$. Lacking a
better semiclassical theory we fit $\langle g\rangle$ to our data, and we do
so in the vicinity of the Heisenberg time. For the fitting we have actually
used the fact that (\ref{facff}) as shown in Fig.~\ref{sb:scl} represents our
data up to and slightly beyond the Heisenberg time very well. $\<g\>$ was
determined as the value for which the long-time expression for the form factor
matches smoothly to this ansatz. We prefer this procedure to a standard
least-square fit over a large time interval, since it emphasizes that the
value of $\<g\>$ is at least implicitly contained in the {\em short-time}
behaviour of the form factor.

With $\langle g\rangle$ obtained in this way, (\protect\ref{fftundis})
describes $K_{0/1}(\tau\ge 1)$ very well (Fig.~\ref{sb:scl}). Since the sum of
the two form factors $K_{0}+K_{1}$ is constant according to (\ref{tc-rmt}) and
Fig.~\ref{sb:scl}b, all information is contained in the difference of the form
factors which is shown in Fig.~\ref{sb:long}b. We regard the good agreement
over a very long time as numerical evidence in favour of the presented theory
for the long-time behavior of the form factor although it contains $\<g\>$ as
a fit parameter.

\subsection{Quantum graphs}
\label{ssec:qgr}
In this subsection we construct and investigate a two-cell system
consisting of a quantized graph. It was recently shown \cite{kot} that
quantum graphs exhibit the common quantum signatures of chaos and
allow for a formally semiclassical description on the basis of
non-deterministic classical dynamics.

\begin{figure}
\centerline{
\psfig{figure=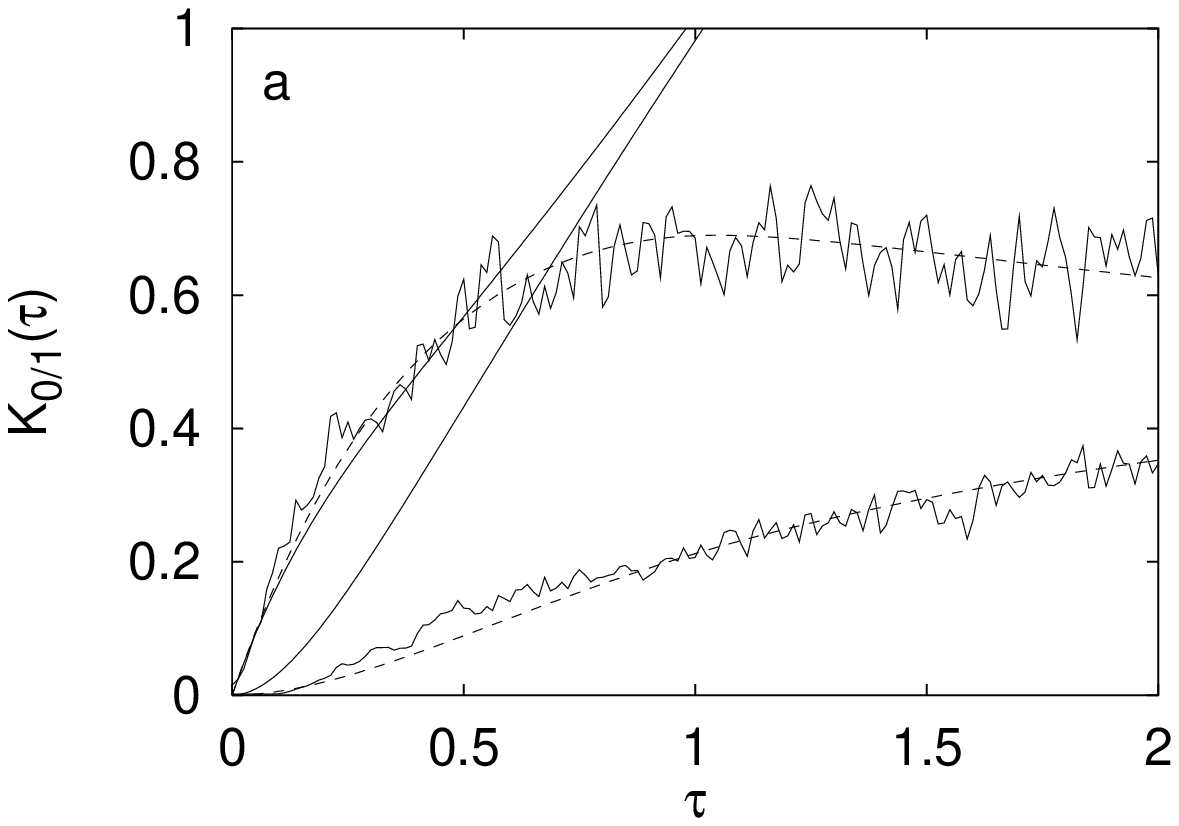,width=6cm}
\psfig{figure=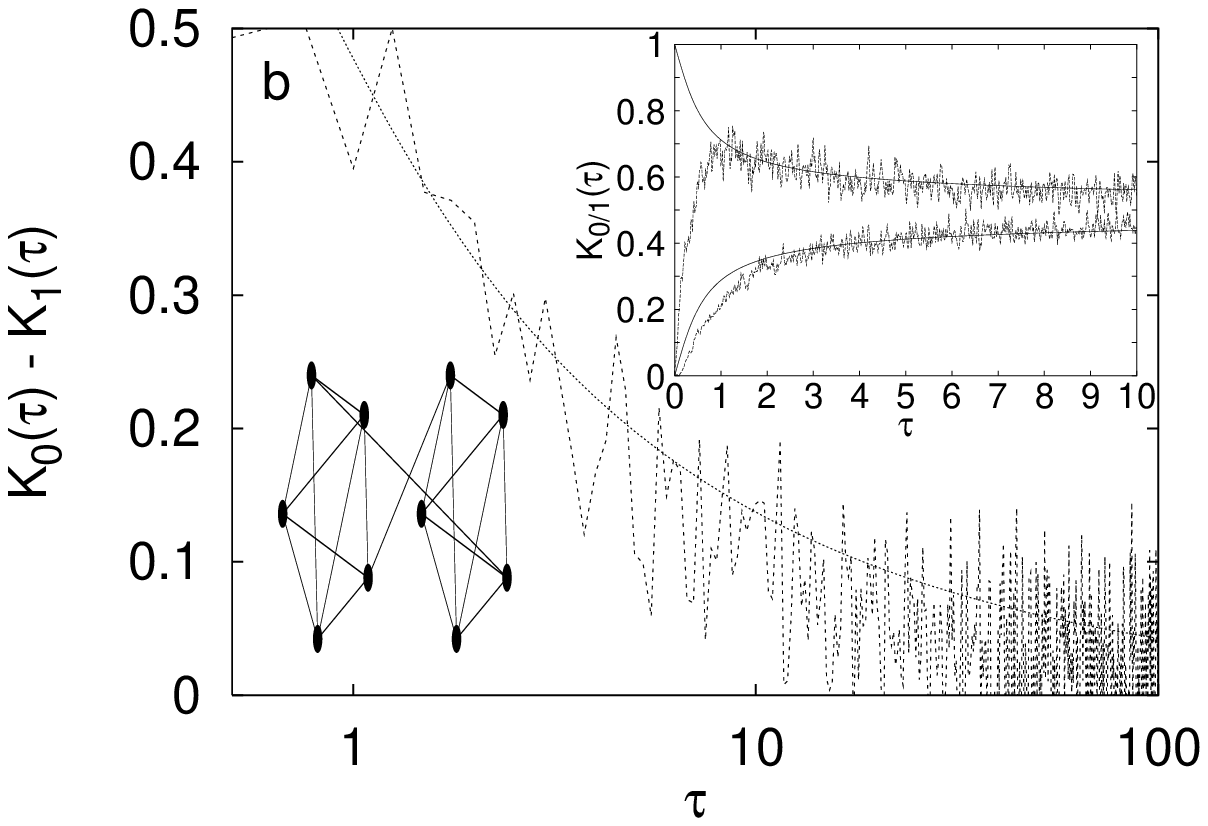,width=6cm}
}
\caption{\label{gr:cmp} (a) Form factors $K_{0/1}(\tau)$ for the two-cell
quantum graph compared to the semiclassical diagonal approximation and the
ansatz (\protect\ref{facff}) as in Fig.~\protect\ref{sb:scl}a.  (b) compares
the difference of the form factors to the theory for the long-time behavior
with the paramter $\<g\>$ obtained as in Fig.~\protect\ref{sb:long}.  The
insets show the analogue of Fig.~\protect\ref{sb:long}a and the topology of
the unit cell.}
\end{figure}

A graph is defined by $v=1,\dots,V$ vertices and $2B$ directed bonds
connecting them. The bond $b$ with length $L_{b}$ is understood to lead from
vertex $v(b)$ to $v(\bar b)$, $\bar b$ being the reversed bond ($L_{\bar
b}=L_{b}$). On each bond we use a coordinate $x_{b}$ with $x_{b}=0$ at $v(b)$,
$x_{b}=L_{b}$ at $v(\bar b)$ and $x_{\bar b}=L_{b}-x_{b}$.  The wave function
$\phi_{b}(x_b)$ satisfies the Schr\"odinger equation ($\hbar=2m=1$)
\begin{equation}\label{se}
\(\[{{\rm d}\/{\rm d}x_b}\]^2+k^2\)\phi_{b}(x_b)=0\,.
\end{equation}
At the vertices, boundary conditions are chosen such that the current is
conserved and the resulting Hamiltonian is self-adjoint and time-reversal
invariant. Following the definitions in \cite{kot}, we require (i), that the
wave function is continuous across all vertices, i.e., it has the same value
in all bonds $b$ connected to some vertex $v$
\begin{equation}\label{cont}
\phi_{b}(0)=\psi_{v=v(b)}\,,
\end{equation}
and (ii), that the sum of the momenta in these bonds vanishes
\begin{equation}\label{bc}
\sum_b\delta_{v,v(b)}{{\rm d}\/{\rm d}x_b}\phi_{b}(0)=0\,.
\end{equation}
The eigenvalues of the so-defined graphs can easily be found numerically
\cite{kot}. For the unit cell, we have interconnected $V=10$ vertices using
$B=20$ bonds such that each vertex is the intersection of exactly four bonds
(inset of Fig.~\ref{gr:cmp}b).  In this case the classical dynamics is
particularly simple: On the bonds, there is free motion at speed $2k$ and each
vertex scatters the particle into any attached bond with equal probability
$1/4$. On basis of such classical dynamics, a formally semiclassical
quantization can be formulated which turns out to be exact in this model.

One of the bonds connecting the two pentagonal layers in Fig.~\ref{gr:cmp}b
was sectioned, and both ends connected to a second identical unit cell, such
that both cells form a ring with translation invariance in the direction
``normal'' to the pentagonal layers. The bond lengths $L_{b}$ of the unit cell
are chosen as random numbers, such that the reflection symmetry is broken.  The
total length is normalized according to $\sum_{b}L_{b=1}^{B}=\pi$ so that the
mean level spacing in $k$ of the unit cell is unity. Therefore it is
advantageous to use the wavenumber $k$ and the path length $L$, instead of
energy and time, as conjugate variables for the semiclassical description. The
Heisenberg time is thus replaced by the Heisenberg length $L_{\rm H}=2\pi$ and
dimensionless time is introduced as $\tau=L/L_{\rm H}$. In the ergodic regime,
the escape rate from the unit cell (again with respect to unit path length
instead of unit time) is simply given by the inverse total length of the graph,
$\lambda=1/\pi$.

It is a particularly favourable feature of the model that neither this rate nor the
number of quantum channels connecting the two cells depend on energy.
Therefore, we can average over arbitrarily large energy intervals. We have
computed the form factor from the 10,000 lowest doublets after dividing the
spectrum into groups of 40 doublets each. Fig.~\ref{gr:cmp} compares the data
as in the previous section to the diagonal approximation, the ansatz
(\ref{facff}) and the long-time theory (\ref{p_d_tau}). The results correspond
to those for the Sinai billiard, but due the larger amount of data the
agreement with our theoretical predictions is even closer.

\subsection{The quantum kicked rotor on a torus}
\label{ssec:qkr}
The kicked rotor belongs to the class of one-dimensional systems that are
rendered classically chaotic only by a periodic driving. Its phase space is
spanned by an angle and an angular-momentum variable and therefore has the
topology of a cylinder. The nonlinearity of the potential is restricted to its
time-dependent component and is controlled by a perturbation parameter.
Accordingly, the classical dynamics crosses over smoothly from integrability
to global chaos with increasing nonlinearity parameter, thereby following the
KAM scenario \cite{lic,chi}. The phase space of the kicked rotor is periodic
also with respect to its non-cyclic coordinate, namely along the
angular-momentum axis.

Quantum-mechanically, the classical angular-mo\-men\-tum period coexists with
$\hbar$ as a second independent action scale. If both are commensurable, then
also the quantum kicked rotor is periodic with respect to angular momentum and
can serve as a model for solid-state-like systems with discrete spatial
translation invariance. Since in the periodic case, the cylindrical phase
space may be regarded as being bent back to itself, this variant of the model
is referred to as the `kicked rotor on a torus'. It is this case which we
shall discuss below. If the two angular-momentum periods are incommensurate,
the quantum eigenstates are generally localized. In this case, the kicked
rotor provides a model for Anderson localization in disordered systems
\cite{dit3,izr}. We will not consider it here.

The kicked rotor is defined by its Hamiltonian
\begin{equation}
H(l,\vartheta;t) ={(l-\Lambda)^2\over 2} + V_{\alpha,k}(\vartheta)
\sum_{m=-\infty}^{\infty} \delta(t-m\tau)\,.
\end{equation}
As a consequence of the periodic time dependence, spectrum and eigenstates are
adequately discussed in terms of quasienergies and Floquet states,
respectively. In addition, the kicked rotor may possess two independent
twofold antiunitary symmetries, both resembling time-reversal invariance. In
order to break them in a controlled manner, an angular-momentum shift
$\Lambda$ has been introduced, and the potential is chosen as \cite{blu}
\begin{equation}
V_{\alpha,k}(\vartheta) = k\left[\cos(\alpha\frac{\pi}{2})\cos\,\vartheta +
\frac{1}{2}\sin(\alpha\frac{\pi}{2})\sin\,2\vartheta\right].
\end{equation}
Here, the global prefactor $k$ determines the degree of nonlinearity. The
appropriate classical measure of nonlinearity, however, is the parameter $K =
k\tau$. If $K \gg 1$, chaotic motion prevails and angular momentum diffuses
without restriction by KAM tori.

The ratio of the classical to the quantum period of action is determined by
the parameter $\tau/4\pi$. If it is rational, i.e., if $\tau = 4\pi p/q$, with
$p$, $q$ coprime, a unit cell along the angular-momentum axis arises that
accomodates $q$ quanta of angular momentum. The number of quasienergy levels
per unit cell is then also $q$. We set $p = 1$ and, in order to avoid an
unwanted symmetry of the unit cell, require $q$ to be odd.

According to Bloch's theorem, the spatial periodicity implies the existence of
an additional constant of the motion, the Bloch phase $\theta$. It appears
explicitly in the symmetry-projected Floquet operator \cite{izr},
\begin{eqnarray}
&& \<\,l'\,|\widehat U(\theta)|\,l\,\> =
\exp\left(-2\pi{\rm i}\frac{p}{q} (l-\Lambda)^2\right) \nonumber\\
&&\quad \times \frac{1}{q} \sum_{n=0}^{q-1}
{\rm e}^{-{\rm i}V_{\alpha,k}([\theta+2\pi n]/q)}
{\rm e}^{{\rm i}(l-l')(\theta+2\pi n)/q}.
\end{eqnarray}
A restriction of the lattice to a finite number of $N$ unit cells, with cyclic
boundary conditions at the ends, amounts to discretizing the Brillouin zone so
that it comprises $N$ equidistant values $\theta_m = 2\pi m/N$, $m = 0$,
$\ldots$, $N-1$, of the Bloch phase. The independent parameter $N$ corresponds
to the number of levels per band, the total number of levels in the spectrum
is $Nq$.

A system with two unit cells is constructed simply by setting $N = 2$. In
contrast to the systems discussed above, the resulting model does not possess
a bottleneck between its two compartments, neither in configuration space nor
in phase space. A reduced exchange between them therefore comes about solely
by slow diffusion. The exchange rate $\lambda$ is determined by the diffusion
constant $D = k^2/2$ (valid if $K \gg 1$) through the simple relation $\lambda
= D/(2a^2)$ derived in \ref{app:dif}. In the quantum kicked rotor
on a torus, the integer $q$ represents the dimensionless size of the
unit-cell, to be substituted for $a$.

The most interesting parameter regime to be studied numerically would be one
where $K_0(\tau)$ exhibits a positive peak around $\tau = 1$, the feature
indicating quasidegeneracy in the spectrum. For this peak to emerge, the
exchange between the cells must be slow. Since there is no bottleneck in the
kicked rotor, this can only be achieved through a small diffusion constant.
More precisely, it requires that the Heisenberg time should be small against
the Thouless time. Measured in units of the discrete time steps of the kicked
rotor they are, respectively, $n_{\rm H} = q$ and $n_{\rm D} = N^2 q^2/(\pi
D)$, so that the condition for quasidegeneracy to occur reads
\begin{equation}
\label{bands}
N^2 \gg \frac{\pi k^2}{2q}.
\end{equation}
At the same time, it should be avoided that localization becomes effective
even within the unit cells, in order to separate the signature of classical
diffusion from the direct quantal effect of disorder in the spectrum. The
localization length should therefore be kept large compared to the size of the
unit cell,
\begin{equation}
\label{noloc}
\xi \approx \frac{k^2}{4} \gg q.
\end{equation}
Clearly, both conditions, (\ref{bands}) and (\ref{noloc}), can hardly be
met simultaneously if $N$ is fixed and small. With $N = 2$, little freedom
remains since, in addition, being close to the classical limit and well within
the classically chaotic regime requires both $q$ and $k$ to be large. We found
that $q = 45$ and $k = 10$ represents an acceptable compromize. The resulting
diffusion constant, corrected for oscillations occurring if $K \appgtr 1$
\cite{lic}, is $D = 23.23$. We substitute the Thouless time $\tau_{\rm D} =
n_{\rm D}/q = 2.467$ for the time constant $1/(\lambda t_{\rm H})$ of the
exponential equilibration.

\begin{figure}
\centerline{
\psfig{figure=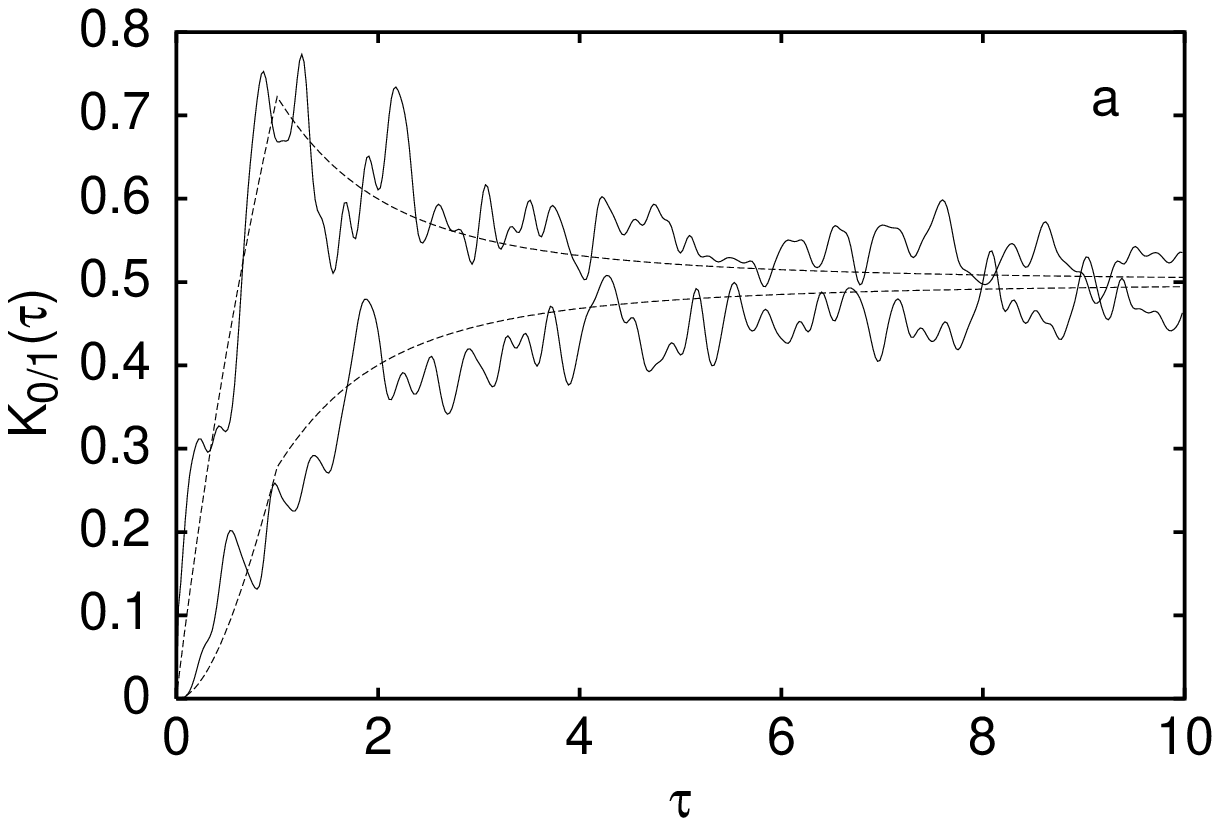,width=6cm}
\psfig{figure=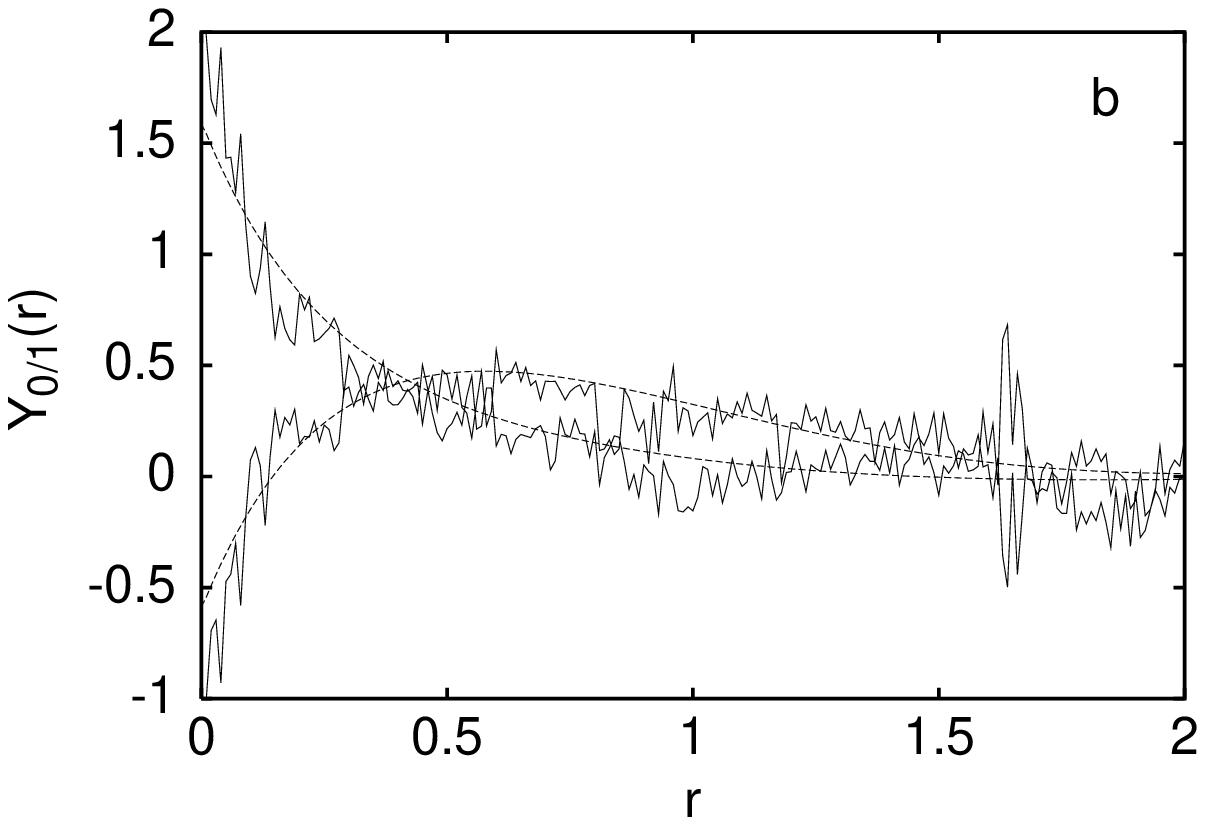,width=6cm}
}
\caption{\label{ffqrt} 
Group-element-specific form factors (a), as described by
Eqs.~(\protect\ref{symffscl}) and (\protect\ref{ffexlongtime}), and
corresponding cluster functions (b), for the quantum kicked rotor on a torus,
compared with the theory according to Eqs.~(\protect\ref{symffscl}) and
(\protect\ref{ffexlongtime}) (dashed). The parameter values are $q = 45$ and
$k = 10$, corresponding to an exchange rate $\lambda = 0.258$.}
\end{figure}
In Fig.~\ref{ffqrt}, we compare the form factors $K_{0/1}(\tau)$ (panel a) and
corresponding cluster functions $Y_{0/1}(r)$ (b) obtained for the quantum
kicked rotor on a torus with parameters as above, with our theory,
Eqs.~(\ref{symffscl}) and (\ref{ffexlongtime}). For the evaluation of the
theory, we have used the relations cited to determine the decay rate directly
from $q$ and $k$. {\em No fitting was involved}. The data cover ten Heisenberg
times and thus reach far into the quantum long-time regime. In the form factors,
we can clearly discern the three time domains discussed, the initial phase of
chaotic diffusion where $K_0(\tau)$ is strongly enhanced while $K_1(\tau)$
remains close to zero, the sharp positive peak of $K_0(\tau)$, reaching almost
twice the asymptotic value, and the saturation regime where $K_0(\tau)$ and
$K_1(\tau)$ approach their common asymptote from above and below,
respectively. The cluster function for $r \applss 1$ represents the regime of
long times or small splittings in a different manner. Both plots give evidence
that the theory provides a quantitative description of the two-point
correlations over all time/energy scales.

\subsection{A random-matrix model}
In the present subsection, we consider
a simple model for a reflection-symmetric
double well system with chaotic dynamics,
with and without time-reversal symmetry,
constructed in the spirit of Refs.~\cite{boh3,tom}.
It will be shown that in certain cases,
this model qualitatively reproduces the features of the form factor discussed
in Sections \ref{sec:scl} and \ref{sec:qum}.

We consider a Hamiltonian of the form
\begin{equation}
\label{eq:model}
H = \left(
\begin{array}{ll}
H_0 & V\\
V & H_0
\end{array}
\right)
\end{equation}
where $H_0$ represents the internal dynamics
of either cell in an $N$-dimensional Hilbert space,
and $V$ their coupling via $M \ll N$ channels of the
connecting duct. Note that in contrast to the models discussed in
\cite{ric,boh3,tom}, we require the two blocks on the diagonal to be identical.

We model $H_0$ as an $N\times N$ random matrix distributed according to Dyson's
Gaussian ensembles, $P(H_0)\,{\rm d}H_{0} \propto \exp(-\mbox{Tr}\,H_0^2/4)\,
{\rm d}H_{0}$. It is assumed that $N\rightarrow\infty$.  The $N\times N$ matrix
$V$ has the form
\begin{equation}
V_{kl} = \delta_{kl} \frac{N}{M} \frac{v\Delta}{\pi^2}
\hspace*{1cm}\mbox{for $k=1,\ldots,M$}
\end{equation}
and zero for $k > M$. Here, $M \ll N$ is the number of matrix elements coupling
the two wells, $v$ parameterizes their strength, and $\Delta$ is the mean level
spacing of $H_0$, $\Delta = \pi\sqrt{\beta/N}$ with $\beta=1$ in the Gaussian
Orthogonal Ensemble (GOE) and $\beta=2$ in the Gaussian Unitary Ensemble
(GUE).

The Hamiltonian $H$ has a twofold symmetry. Its eigenvalues can be classified
according to parity $p$ and appear as doublets $r_{\alpha,\nu}$ with
$\alpha=1,\ldots,N$ and $\nu=\pm$. According to Eq.~(\ref{doublets}), we write
$r_{\alpha,\pm} = R_\alpha \pm r_\alpha$. The form factor is then given by (cf.\
Eq.~(\ref{fftunsplit}))
\begin{equation}
\label{eq:ktstart}
K_{0/1}(\tau)  = 
\frac{1}{2} \pm \left\langle\frac{1}{2N} \sum_{\alpha=1}^N \cos(4\pi \tau
r_\alpha) \right \rangle\,.
\end{equation}
For large times ($\tau \gg 1$), $K_{0/1}(\tau)$ may be calculated by
evaluating the doublet splitting $2 r_\alpha$ within degenerate perturbation
theory. Denoting the eigenfunctions of $H_0$ by $\phi_\alpha$ (with
components $\phi_{\alpha\nu}$), one has 
(with $\langle d_{\rm fd}\rangle = \Delta^{-1}$)
\begin{equation}
\label{eq:pt}
r_\alpha \simeq \frac{v}{\pi^2}\frac{N}{M} \sum_{\nu=1}^M
|\phi_{\alpha\nu}|^2\,.
\end{equation}
Substituting (\ref{eq:pt}) into (\ref{eq:ktstart}), it remains to average over
the eigenfunctions $\phi_\alpha$ of $H_0$. The statistical properties of the
eigenfunctions $\phi_\alpha$ depend on the ensemble considered. In the GUE,
the amplitude $u = N|\phi_{\alpha\nu}|^2$ is distributed according to $P(u) =
\exp(-u)$. In the GOE, the corresponding distribution is 
$P(u) = (2\pi u)^{-1/2} \exp(-u/2)$.

We first consider the case $M=1$, where the two wells are coupled via a single
matrix element. For $\tau \gg 1$, one obtains for $K_{0/1}(\tau)$
\begin{equation}
\label{eq:N1}
\nonumber
K_{0/1}(\tau)
\approx
\frac{1}{2} \pm
\frac{1}{2}
\left \{
\begin{array}{ll}
\frac{1}{4}\left({v\tau/\pi}\right)^{-1/2} & \mbox{for $\beta = 1$}\\[0.2cm]
1/[1+(4v\tau/\pi)^2] &\mbox{for $\beta = 2$}.
\end{array}
\right .
\end{equation}
For $\beta=1$, this expression reproduces the long-time $\tau^{-1/2}$ decay of
Eq.~(\ref{asym-trs}). For $\beta=2$, Eq.~(\ref{eq:N1}) reproduces the
$\tau^{-2}$ decay for large $\tau$ implied by Eq.~(\ref{uncondexft}).

\begin{figure}
\centerline{\psfig{file=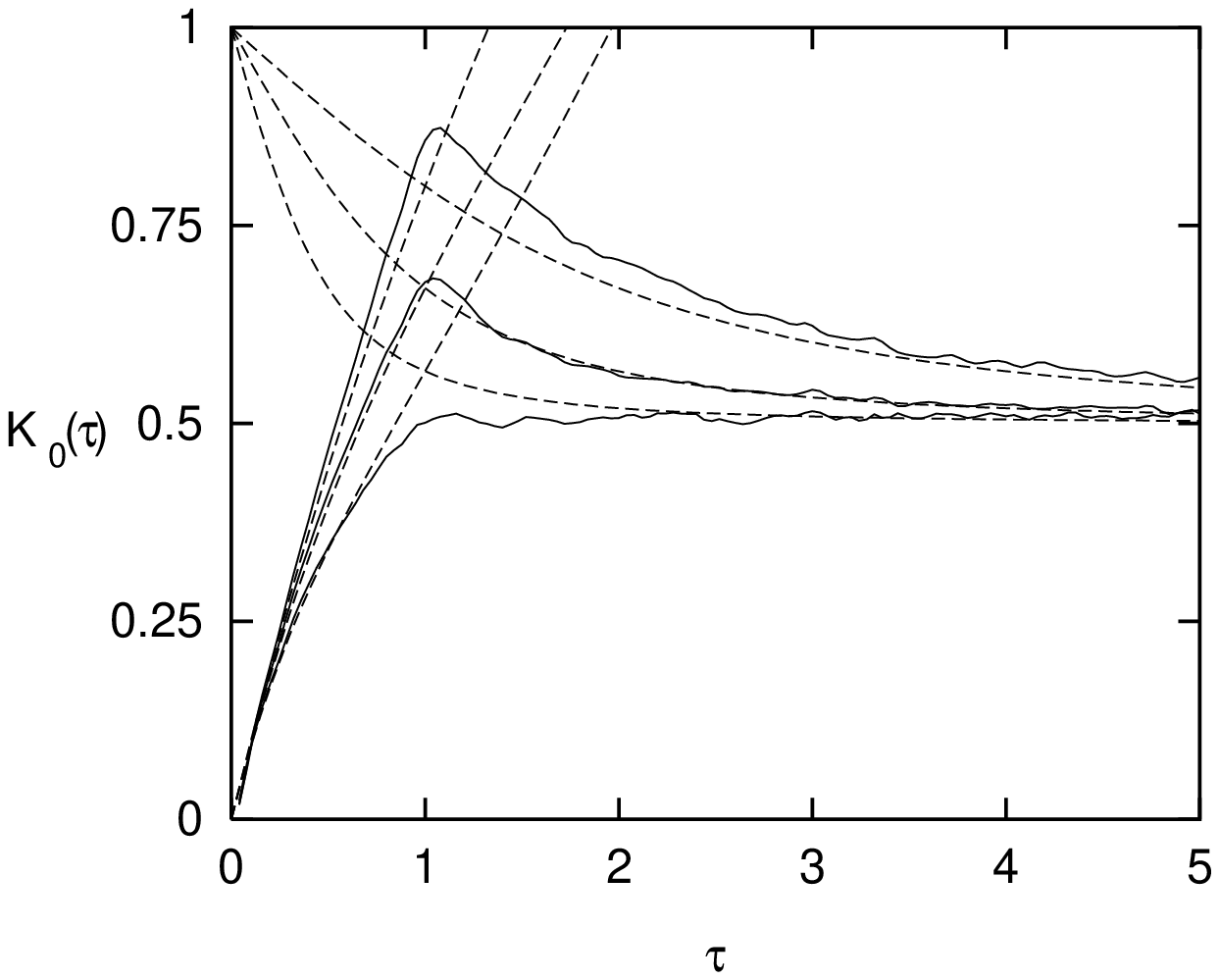,width=6cm,angle=270}
            \psfig{file=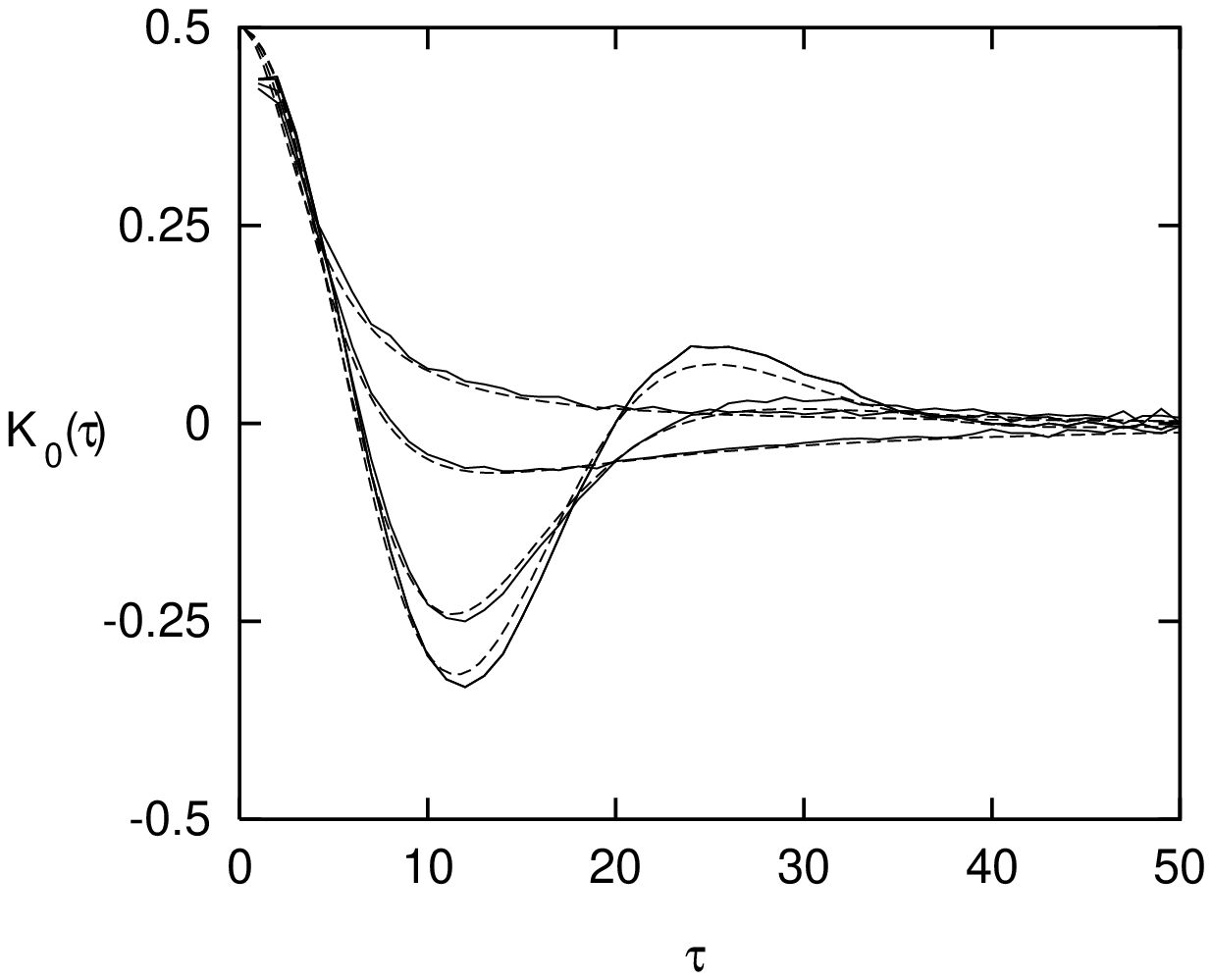,width=6cm,angle=270}}
\caption{\label{fig:kt} (a) 
Form factor $K_0(\tau)$ in the GUE
for three different values of the coupling strength ($v=0.5$, 1, and 2), and for
$N=80$ and $M=1$ (full lines). Also shown are the theoretical results,
Eqs.~(\protect\ref{symffscl}) and (\protect\ref{ffexlongtime}) (dashed).
(b) Form factor $K_0(\tau)-1/2$ in the GUE
for $v=0.2$, $N=80$ and $M=1$, 2, 6, and 10 (full lines),
compared to the asymptotic theoretical result,
Eq.~(\protect\ref{eq:Nlarge}) (dashed).
}
\end{figure}

For $\beta = 2$, the matching procedure
discussed in Section \ref{sec:qum} yields
an analytical expression for $K_0(\tau)$
valid for all time scales. In Fig.~\ref{fig:kt}a,
we compare this expression (Eqs.~(\ref{symffscl}), (\ref{ffexlongtime}))
with results of simulations of
the model (\ref{eq:model}). Shown
is $K_0(\tau)$ as a function of $\tau$
for an ensemble of random matrices with $N=80$,
$v=0.5,1$ and $2$ in the GUE (full lines),
as well as Eqs.~(\ref{symffscl}) and (\ref{ffexlongtime}).
The constant $v$ may be determined
from $\langle g\rangle$ by comparison of Eq.~(\ref{eq:N1}) with
Eq.~(\ref{ffexlongtime}),
where $c = 8\pi \langle g\rangle$.

We find good agreement between
the results of the simulations
and Eqs.~(\ref{symffscl}), (\ref{ffexlongtime}). We have,
however, not attempted to
evaluate the small-$\tau$ behaviour
of the form factor for the model
(\ref{eq:model}) analytically. For large $\tau$,
on the other hand, it is clear
that Eq.~(\ref{eq:model}) is a good model for
the form factor: as pointed out above,
Eqs.~(\ref{eq:N1}) and (\ref{ffexlongtime}) coincide for large $\tau$.

In the present model, it is also possible to consider larger $M$,
$1 < M \ll N$. In this case, for $\beta = 2$, the quantity
\begin{equation}
u = \frac{N}{M} \sum_{\nu=1}^M |\phi_{\alpha\nu}|^2
\end{equation}
is distributed according to
\begin{equation}
P(u,M) = \frac{M^M u^{M-1} {\rm e}^{-M u}}{\Gamma(M)}\,.
\end{equation}
We require $M \ll N$ since for $M=N$, one has $u=1$ due to normalization of the
wave functions. In the GUE, the form factor is then given by
\begin{eqnarray}
\nonumber
K_0(\tau) &\approx& \frac{1}{2}+ \frac{1}{2}\int_0^\infty\!
{\rm d} u\, \cos\Big(\frac{4v}{\pi} \tau u\Big)\, P(u,M)\\
\label{eq:Nlarge}
& = & \frac{1}{2}
+ \frac{1}{2}\left(1+\Big(\frac{4v\tau}{\pi M}\Big)^2\right)^{-M/2}\,
\cos\left(M\mbox{atan}\left(\frac{4v\tau}{\pi M}\right)\right)\,.
\end{eqnarray} For $M=1$, Eq.~(\ref{eq:N1}) is reproduced.
For $1 < M \ll N$, the form factor does not decay monotonically for $\tau > 1$
but exhibits oscillatory behaviour (Fig. \ref{fig:kt}b).

\section{Conclusion}
\label{sec:sum}
Chaotic systems with two weakly connected cells, elementary as this concept
may appear, form a paradigm for a large class of physical situations and
exhibit a surprisingly rich behaviour. In this paper, we have shown that it is
determined essentially by two parameters. One of them can be identified with the
time required for the respective populations of the cells to equilibrate. It
specifies the position between the extreme of a large opening that hardly
restricts the exchange, and the opposite one of two almost isolated single cells.
The second relevant parameter is a measure of the difference in shape between
the cells, ranging from exact symmetry to its complete absence.

On basis of the results obtained in this paper, we can draw a clear picture of
the spectral two-point correlations in this two-dimensional parameter space.
In the case of an effective communication between the cells, the presence or
absence of symmetry is of little relevance for the spectral statistics. It is
then only the slight retardation of ergodic coverage that becomes manifest in
the level correlations. The result is a reduction in the area enclosed by the
initial minimum (correlation hole) of the form factor, indicating an increase
of randomness in the spectrum which can be completely accounted for by
semiclassical considerations \cite{smi}.

The case of two almost isolated cells lacking all symmetry can be trivially
understood from a random-matrix point of view. We are then dealing with the
superposition of two spectra that are nearly mutually independent but exhibit
the same statistics. Here, random-matrix theory simply predicts a doubling of
the time argument of the form factor \cite{boh2}, in agreement with the
semiclassical approximation in the limit of slow equilibration.

If, in contrast, the two cells are symmetric, the formation of doublets
introduces an additional feature in the spectrum. The corresponding positive
correlations are reflected in the form factor as a maximum in the vicinity of
the Heisenberg time. In the limit of long exchange time, the form factor at
this maximum reaches twice the asymptotic value to which it decays
subsequently from above, relative to its value at $t = 0$. In the case of
exact symmetry, we can quantitatively account for this peak in the
standard form factor. Simultaneously, there is a depression in an
analogous statistic that refers to transport from one cell to the other, rather
than to return to the initial one.

The crossover from full to completely broken symmetry, as a function of some
symmetry-breaking parameter, can be included in the semiclassical theory if a few
plausible additional assumptions are made. In accordance with corresponding work
on spatially periodic systems with slight disorder \cite{dit4}, this approach
implies that the peak in the form factor should decay exponentially both with
the typical difference in action between symmetry-related periodic orbits in the
respective cells, and with time.

\ack One of us (TD) would like to thank for the warm hospitality enjoyed
during a stay in the group of Prof.~O.~Bohigas, Institut de Physique
Nucl\'eaire, Orsay. This stay was financed by Deutsche Forschungsgemeinschaft
(grant No.\ Di~511/4-1). BM gratefully acknowledges financial support by the
{\em Sonderforschungsbereich 393 ``Numerische Simulation auf massiv parallelen
Rechnern''} of the DFG.

\appendix

\section{Discrete diffusion}
\label{app:dif}
The picture of a two-cell system with the topology of a ring suggests to
unroll the ring so that an infinite chain is formed. If the two cells are
translation symmetric, then each of them represents a unit cell of this
periodic lattice, otherwise the unit cell comprises both cells of the ring. In
this extended topology, the condition of slow exchange between the cells
implies that the picture of homogeneous diffusion breaks down on the scale of
the lattice constant. We are therefore in a regime opposite to that considered
in Ref.~\cite{dit1}.

The spreading along the chain is now determined by a master equation for the
probability to be at site $n$, instead of a diffusion equation for the
probability density,
\begin{equation}
\dot{P}_n(t) = \lambda P_{n-1}(t) - 2\lambda P_n(t) + \lambda P_{n+1}(t),
\end{equation}
In a more concise notation, it reads
\begin{equation}
\left({{\rm d}\over{\rm d}t} - \lambda\Delta_2\right){\bm P}(t) = 0.
\end{equation}
Here, ${\bm P}(t)$ denotes the entire infinite vector of the $P_n(t)$, and
\begin{equation}
\Delta_2 = \left(
\begin{array} {ccccccccc}
.&.&.&  &  &  & & & \\
 &.&.& 1&  &  & &0& \\
 & &1&-2& 1&  & & & \\
 & & & 1&-2& 1& & & \\
 & & &  & 1&-2&1& & \\
 &0& &  &  & 1&.&.& \\
 & & &  &  &  &.&.&.
\end{array} \right)
\end{equation}
is the discrete Laplace operator. The lattice plane waves
\begin{equation}
\phi_{m,n} = \frac{1}{\sqrt{N}}{\rm e}^{2\pi{\rm i}nm/N},\quad
m = 0,\ldots,N-1,
\end{equation}
where for the sake of normalizability we have reintroduced cyclic boundary
conditions with a period of $N$ chain elements, solve the stationary
eigenvalue equation
\begin{equation}
\Delta_2 {\bm\phi}_m = \gamma_m {\bm\phi}_m
\end{equation}
with eigenvalues
\begin{equation}
\gamma_m = 2\left(\cos\frac{2\pi m}{N} - 1\right).
\end{equation}
For a localized initial state
\begin{equation}
P_n(0) = \delta_{n\,{\rm mod}\,N} =
\frac{1}{\sqrt{N}}\sum_{m=0}^{N-1} \phi_{m,n}\,,
\end{equation}
the time evolution reads
\begin{eqnarray}
\label{disdiffsol}
P_n(t) &=& \frac{1}{\sqrt{N}} \sum_{m=0}^{N-1}
\phi_{m,n} {\rm e}^{\gamma_m t}\nonumber\\
&=& \frac{1}{N} \sum_{m=0}^{N-1}
\exp\left(2\pi{\rm i}\frac{mn}{N}+
2\lambda\left[\cos\frac{2\pi m}{N} -
1\right]t\right).
\end{eqnarray}
By Poisson resummation, this becomes
\begin{eqnarray}
P_n(t) &=& \frac{1}{N} \sum_{m=-\infty}^{\infty} \int_0^N{\rm d}\nu\,
\exp\left(2\pi{\rm i}\left[m+\frac{n}{N}\right]\nu+
2\lambda\left[\cos\frac{2\pi \nu}{N} - 1\right]t\right)\nonumber\\
&=& {\rm e}^{-2\lambda t}
\sum_{m=-\infty}^{\infty} {\rm I}_{|n+mN|}(2\lambda t),
\end{eqnarray}
where ${\rm I}_n(z)$ denotes the modified Bessel function of integer order $n$
\cite{abr1}. The spreading over the lattice, as described by $P_n(t)$,
represents a discrete diffusion process. If we go to the
continuum limit by defining $x = na$, $L = Na$, and letting the lattice
constant $a \to 0$, we recover continuous diffusion with periodic
boundary conditions,
\begin{equation}
p(x,t) = \frac{1}{a} P_n(t) \to \frac{1}{\sqrt{2\pi Dt}}
\sum_{m=-\infty}^{\infty} \exp\left(-\frac{(x+mL)^2}{2Dt}\right).
\end{equation}
The diffusion constant is $D = 2\lambda a^2$. In performing the limit, we have
used the asymptotic form of the ${\rm I}_n(z)$ for large argument $z$
\cite{abr2} and expanded it for large order $n$. The two-cell solution,
Eq.~(\ref{mastersol}), is retained by setting $N = 2$ in
Eq.~(\ref{disdiffsol}).

\section{Doublets as discretized bands}
\label{app:ban}
As on the level of the classical dynamics, it is instructive to consider also
the quantum two-cell system as the unit cell of an infinite chain. From this
point of view, the doublets $r_{\alpha,\pm}$ come about by discretizing
continuous bands to a ``Brillouin zone'' with only two points. The simplest
possible interpolation between these points assumes cosine-shaped bands,
\begin{equation}
\label{eq:cosbands}
r_{\alpha}(\mu) = R_{\alpha} + r_{\alpha} \cos(\pi\mu),\quad
\mu = 0,1.
\end{equation}
Equation (\ref{eq:cosbands}) can be justified by the fact that it imposes no
more information on the shape of the bands than is available, namely their
first two Fourier coefficients. Cosine-shaped bands result also from
diagonalizing a tight-binding Hamiltonian with translation invariance. For two
sites this is
\begin{eqnarray}
H_{n,n'}^{(\alpha)} =&&
R_{\alpha}\delta_{(n-n')\,{\rm mod}\,2} \nonumber\\
&& + \frac{1}{2} r_{\alpha}
(\delta_{(n-n'-1)\,{\rm mod}\,2} +
\delta_{(n-n'+1)\,{\rm mod}\,2}), \nonumber\\
&&\hspace*{4cm} n,n' = 0,1.
\end{eqnarray}
We have defined the parameters of this Hamiltonian in such a way that
Eq.~(\ref{eq:cosbands}) gives its eigenenergies. Inserting them in
Eq.~(\ref{twoelamp}) and performing a Poisson resummation results in
\begin{equation}
\label{poissonres}
a_n(\tau) = \sum_{m=-\infty}^{\infty} \sum_{\alpha = 1}^{N_{\rm d}}
{\rm e}^{-2\pi{\rm i}\tau R_{\alpha}} \, {\rm i}^{n-2m} \,
{\rm J}_{2m-n}(2\pi\tau r_{\alpha}),
\end{equation}
where ${\rm J}_k(z)$ denotes the ordinary Bessel function of order $k$.
We introduce a diagonal approximation with respect to the band index $\alpha$,
as in Section \ref{sec:qum}, and obtain the corresponding form factors as
\begin{equation}
K_n(\tau) = \frac{1}{N_{\rm d}} \sum_{\alpha = 1}^{N_{\rm d}}
\left|\sum_{m=-\infty}^{\infty} (-1)^m
{\rm J}_{2m-n}(2\pi\tau r_{\alpha})\right|^2.
\end{equation}
Invoking the sum rules $\sum_{k=-\infty}^{\infty} (-1)^k {\rm J}_{2k}(z) =
\cos z$, $\sum_{k=-\infty}^{\infty} (-1)^k {\rm J}_{2k-1}(z) = \sin z$
\cite{abr3}, we recover Eq.~(\ref{fftunsplit}).

\section{Quantization of the Z-shaped billiard}
\label{sssec:qm}

Quantization of a billiard amounts to solving the Helmholtz equation
\begin{equation}
\left(\frac{\partial^2}{\partial x^2} + \frac{\partial^2}{\partial y^2} + 
k^2\right)\psi(x,y)=0
\end{equation}
with Dirichlet boundary conditions on the billiard circumference and a
dispersion $k^2=2mE/\hbar^2$.  In this appendix we describe a specific
quantization method for the Z-shaped billiard discussed in Section
\ref{ssec:sinai1}.  It is based on the scattering approach
\cite{dor,sch}. Consider a subdivision of the closed double billiard into two
open halves (Fig.~\ref{hdb-fig}). Each of them represents a chaotic scatterer
attached to the end of a semi-infinite waveguide. Within the waveguide,
quantization of transverse momentum, $k_{y,n} = n\pi$, $n = 0$, $\pm 1$, $\pm
2$, $\ldots$, implies that there are $N = [k/\pi]$ ($[\dots]$ denoting integer
part) open channels with real longitudinal momentum $k_{x,n} = (k^2-
k_{y,n}^2)^{1/2}$, such that the two scatterers are described by
$N\,\times\,N$ scattering matrices $S^{\rm l}$ and $S^{\rm r}$, respectively.

The secular equation for the eigenvalues of the full billiard then reads
\begin{equation}
\det(I-S^{\rm l}(k)S^{\rm r}(k))=0.
\end{equation}
In order to construct the $S^{\rm l/r}$ for the billiard halves
\cite{dor}, we start from the $2N\,\times\,2N$ transfer matrix for a
quarter Sinai billiard open on both sides \cite{dit2,dit3},
\begin{equation}
T^{\rm s} = \left(\begin{array}{cc} rt^{-1}r-t & rt^{-1}\\
           t^{-1}r & t^{-1}  \end{array}\right)\; ,
\end{equation}
Here, $t$ and $r$ denote the $N\,\times\,N$ matrices of transmission and
reflection amplitudes, respectively. Due to the spatial reflection symmetry
with respect to the diagonal, the two entrances of the billiard are
equivalent.

The transfer matrix $T^{\rm w}$ for a waveguide of length $L/2$ consists of
phase factors $\exp(\pm{\rm i}k_{x,n}L/2)$ along the
diagonal. The letter-Z-like fashion in which the two halves are assembled is
accounted for by a third factor $T^{\rm z}$ with appropriate phases $\pm 1$
\cite{dit2,dit3} along its diagonal. It is included in the transfer matrix for
one of the sides, e.g., $T^{\rm l} = T^{\rm e}T^{\rm w}$, $T^{\rm r} = T^{\rm
l}T^{\rm z}$.

The scattering matrices for the billiard halves closed on one side are
obtained from $T^{\rm l}$ and $T^{\rm r}$ by requiring incoming and outgoing
amplitudes to cancel across the openings where Dirichlet boundary conditions
are to be enforced,
\begin{equation}
\left(\begin{array}{c}{+\bm A}\\-{\bm A}\end{array}\right)=
T^{\rm l/r}\left(\begin{array}{c}{\bm B}\\{\bm C}\end{array}\right).
\end{equation}
Here, ${\bm A}$ and $-{\bm A}$ refer to the amplitudes at the ends to be
closed, and ${\bm B}$ and ${\bm C}$ to the amplitudes on the opposite
sides. The latter are related by ${\bm C} = S^{\rm l/r}{\bm B}$, invoking the
S matrices sought for. Solving for them, one finds
\begin{equation}
S^{\rm l/r}=-(T^{\rm l/r}_{12}+T^{\rm l/r}_{22})^{-1}(T^{\rm l/r}_{11}
              +T^{\rm l/r}_{21}).
\end{equation}
In obvious notation, $T^{\rm l/r}_{ij}$, $i,j = 1,2$, refer to the four
$N\!\times\! N$ subblocks of the respective transfer matrices. 

\begin{figure}
\centerline{\psfig{figure=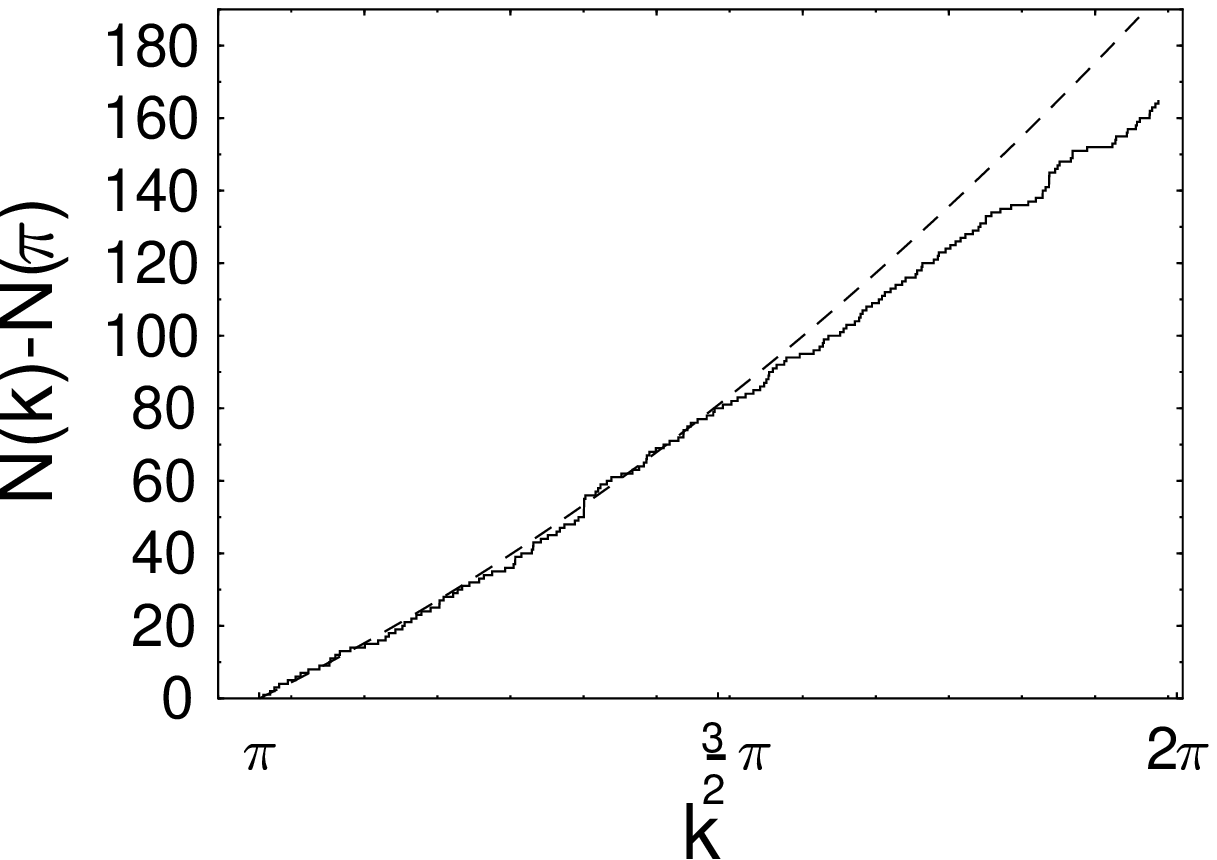,width=7cm}}
\caption{\label{weyl-fig} 
The cumulated eigenvalue density of the double billiard (steps) compared to
the Brownell formula (dashed line). The geometric parameter values are
$R = 10$ and $L = 10$.}
\end{figure}

In order to check the quality of the quantization procedure, we compare, in
Fig.~\ref{weyl-fig}, the numerical result for the cumulated eigenvalue density
with the Brownell formula \cite{bal}, for wavenumbers in the interval $\pi < k
< 2\pi$, and $R = 10$. The agreement is satisfactory up to $k \approx 5$. For
larger wavenumbers, quasidegenerate pairs of zeros occur in the secular
function with too small spacing to be resolved by the numerical
procedure. We have therefore discarded data with $k > 5$. To achieve better
statistics in the evaluation of spectral correlations, we have varied $L$
within an interval $\Delta L$ amounting to a few percent of $L$.

\end{document}